\documentclass[a4paper,fleqn]{cas-dc}

\usepackage[utf8x]{inputenc}

\newif\ifpreprint
\newif\ifshowchanges
\newif\iflineno

\preprinttrue

\ifpreprint
    \linenofalse
\else
    \ifshowchanges
        \linenofalse
    \else
        \linenotrue
    \fi
\fi

\iflineno
    \usepackage[switch,mathlines]{lineno}
    \let\oldequation\equation
    \let\oldendequation\endequation
    
    \renewenvironment{equation}
      {\linenomathNonumbers\oldequation}
      {\oldendequation\endlinenomath}
\else      
    \newenvironment{nolinenumbers}{}{}
\fi

\usepackage[authoryear]{natbib}

\usepackage[thai,main=english]{babel}

\usepackage{tipa}

\usepackage{lipsum}

\usepackage{tabularx}
\usepackage{booktabs}
\usepackage{longtable}
\usepackage{caption}
\usepackage{float}
\usepackage{ragged2e}
\usepackage{caption}

\usepackage{siunitx}
\usepackage{cleveref}

\usepackage{placeins}

\usepackage{etoolbox}
\usepackage{xspace}

\usepackage{savesym}
\savesymbol{comment} % rename \comment -> \origcomment

\ifshowchanges
    \usepackage{changes}
\else
    \usepackage[final]{changes}
\fi

\setcommentmarkup{%
    \marginpar{\small\centering\color{purple}\textbf{[{#1}]}}%
}
\setdeletedmarkup{{\color{red}\sout{#1}}}
\newenvironment{sadded}{\ifshowchanges\color{blue}\fi}{}
\newenvironment{smoved}{\ifshowchanges\color{purple}\fi}{}

\ExplSyntaxOn
\RenewDocumentCommand \eadauthor {} 
    { 
      \seq_map_inline:Nn \l_stm_au_seq 
        { 
            \regex_extract_once:nnNTF {(\w)\w*-(\w)} { ##1 } \l_stm_au_fn_seq
            { 
                \seq_pop_left:NN \l_stm_au_fn_seq \temp_var
                \seq_use:Nn \l_stm_au_fn_seq { .- }
                { . } 
            }
            { 
                \regex_match:nnTF { \. } { ##1 } 
                { ##1 }
                { \tl_head:n {##1}. }
            }
      }{ ~\l_stm_au_sn_seq }
    }
\ExplSyntaxOff
\newif\ifabbreviation
\pretocmd{\thebibliography}{\abbreviationfalse}{}{}
\AtBeginDocument{\abbreviationtrue}
\DeclareRobustCommand\acroauthor[2]{%
  \ifabbreviation
    \ifcsname acroused@#2\endcsname
      #2%
    \else
      #1%
      ~[\mbox{#2}]% <----
      \expandafter\gdef\csname acroused@#2\endcsname{}%
    \fi
  \else
    \ifcsname bibacroused@#2\endcsname
        \mbox{#2}%
    \else
        \mbox{#1}~(\mbox{#2})%
        \expandafter\gdef\csname bibacroused@#2\endcsname{}%
    \fi
  \fi
}

\begin{document}\sloppy
\let\WriteBookmarks\relax
\def\floatpagepagefraction{1}
\def\textpagefraction{.001}
% Thai shorthands (must be loaded after begin document)
\newcommand{\thaiword}[1]{\hspace{-0.5ex}\foreignlanguage{thai}{#1}\xspace}
\newcommand{\lit}[1]{\textit{lit.}\xspace\textit{{#1}}}
\newcommand{\placeholder}{{\color{lightgray}\lipsum/3-4/}}

\newcommand{\naaramkaan}{\textipa{/n\^{a}:.r\={a}m.k\super{h}\={a}:n/}}
\newcommand{\rakaaihuu}{\textipa{/r\'{a}P.k\super{h}\={a}:j.h\v{u}:/}}
\newcommand{\pornklaai}{\textipa{/p\super{h}\`{O}n.k\super{h}l\={a}:j/}}
\newcommand{\sangop}{\textipa{/s\`{a}P.N\`{o}p/}}
\newcommand{\sabaaihuu}{\textipa{/s\`{a}P.b\={a}:j.h\v{u}:/}}
\newcommand{\jorjae}{\textipa{/c\={O}:.c\={E}:/}}
\newcommand{\bpanbpuuan}{\textipa{/p\`{a}n.p\`{u}an/}}
\newcommand{\yungyerng}{\textipa{/j\^{u}N.j\v{7}:N/}}
\newcommand{\wunwaai}{\textipa{/w\^{u}n.w\={a}:j/}}
\newcommand{\miiarai}{\textipa{/m\={i}:.P\`{a}P.r\={a}j/}}
\newcommand{\uekgatuek}{\textipa{/P\`{W}k.k\`{a}P.t\super{h}\'{W}k/}}
\newcommand{\juedjued}{\textipa{/c\`{W}:t.c\`{W}:t/}}
\newcommand{\naabuea}{\textipa{/n\^{a}:.b\`{W}a/}}
\newcommand{\nuuainuuai}{\textipa{/n\={W}aj.n\={W}aj/}}
\newcommand{\uuaiuuai}{\textipa{/P\`{W}aj.P\`{W}aj/}}
\newcommand{\naafang}{\textipa{/n\^{a}:.f\={a}N/}}
\newcommand{\ruenhuu}{\textipa{/r\^{W}:n.h\v{u}:/}}
\newcommand{\sanorhuu}{\textipa{/s\`{a}P.n\`{O}P.h\v{u}:/}}
\newcommand{\riiap}{\textipa{/r\^{i}ap/}}
\newcommand{\maimiiarai}{\textipa{/m\^{a}j.m\={i}:.P\`{a}P.r\={a}j/}}
\newcommand{\kuekkak}{\textipa{/k\super{h}\'{W}k.k\super{h}\'{a}k/}}
\newcommand{\miichiiwitchiiwaa}{\textipa{/mi:.c\super{h}\={i}:.w\'{i}t.c\super{h}\={i}:.w\={a}:/}}
\newcommand{\sotsai}{\textipa{/s\`{o}t.s\v{a}j/}}
\newcommand{\wuewaa}{\textipa{/w\v{W}:.w\v{a}:/}}

\newcommand{\thainaaramkaan}{\thaiword{น่ารำคาญ}}
\newcommand{\thairakaaihuu}{\thaiword{ระคายหู}}
\newcommand{\thaipornklaai}{\thaiword{ผ่อนคลาย}}
\newcommand{\thaisangop}{\thaiword{สงบ}}
\newcommand{\thaisabaaihuu}{\thaiword{สบายหู}}
\newcommand{\thaijorjae}{\thaiword{จอแจ}}
\newcommand{\thaibpanbpuuan}{\thaiword{ปั่นป่วน}}
\newcommand{\thaiyungyerng}{\thaiword{ยุ่งเหยิง}}
\newcommand{\thaiwunwaai}{\thaiword{วุ่นวาย}}
\newcommand{\thaimiiarai}{\thaiword{มีอะไร}}
\newcommand{\thaiuekgatuek}{\thaiword{อึกทึก}}
\newcommand{\thaijuedjued}{\thaiword{จืด ๆ}}
\newcommand{\thainaabuea}{\thaiword{น่าเบื่อ}}
\newcommand{\thainuuainuuai}{\thaiword{เนือย ๆ}}
\newcommand{\thaiuuaiuuai}{\thaiword{เอื่อย ๆ}}
\newcommand{\thainaafang}{\thaiword{น่าฟัง}}
\newcommand{\thairuenhuu}{\thaiword{รื่นหู}}
\newcommand{\thaisanorhuu}{\thaiword{เสนาะหู}}
\newcommand{\thairiiap}{\thaiword{เรียบ}}
\newcommand{\thaimaimiiarai}{\thaiword{ไม่มีอะไร}}
\newcommand{\thaikuekkak}{\thaiword{คึกคัก}}
\newcommand{\thaimiichiiwitchiiwaa}{\thaiword{มีชีวิตชีวา}}
\newcommand{\thaisotsai}{\thaiword{สดใส}}
\newcommand{\thaiwuewaa}{\thaiword{หวือหวา}}

\newcommand{\thaimai}{\thaiword{`ไม่'}}
\newcommand{\mai}{\textipa{/m\^{a}j/}}

\newcommand{\thaiwali}{\thaiword{วลี}}
\newcommand{\wali}{\textipa{/w\'{a}P.l\={i}:/}}

\newcommand{\thaihuu}{\thaiword{`หู'}}
\newcommand{\huu}{\textipa{/h\v{u}:/}}

\newcommand{\thaiarai}{\thaiword{`อะไร'}}
\newcommand{\arai}{\textipa{/P\`{a}P.r\={a}j/}}

\newcommand{\thaimii}{\thaiword{`มี'}}
\newcommand{\mii}{\textipa{/m\={i}:/}}

\newcommand{\thairuen}{\thaiword{`รื่น'}}
\newcommand{\ruen}{\textipa{/r\^{W}:n/}}

\newcommand{\thairakaai}{\thaiword{`ระคาย'}}
\newcommand{\rakaai}{\textipa{/r\'{a}P.k\super{h}\={a}:j/}}

\newcommand{\thaifang}{\thaiword{`ฟัง'}}
\newcommand{\fang}{\textipa{/f\={a}N/}}
\newcommand{\pleasantAPPRkwp}{\num{0.486}}
\newcommand{\pleasantUNDRkwp}{\num{2.88e-07}}
\newcommand{\pleasantCLARkwp}{\num{0.695}}
\newcommand{\pleasantANTOkwp}{\num{0.957}}
\newcommand{\pleasantORTHkwp}{\num{0.206}}
\newcommand{\pleasantNCONkwp}{\num{0.798}}
\newcommand{\pleasantIBALkwp}{\num{0.429}}
\newcommand{\annoyingAPPRkwp}{\num{3.33e-08}}
\newcommand{\annoyingUNDRkwp}{\num{1.65e-07}}
\newcommand{\annoyingCLARkwp}{\num{0.179}}
\newcommand{\annoyingANTOkwp}{\num{0.0854}}
\newcommand{\annoyingORTHkwp}{\num{0.801}}
\newcommand{\annoyingNCONkwp}{\num{0.128}}
\newcommand{\annoyingIBALkwp}{\num{0.859}}
\newcommand{\eventfulAPPRkwp}{\num{0.0296}}
\newcommand{\eventfulUNDRkwp}{\num{1.87e-06}}
\newcommand{\eventfulCLARkwp}{\num{1.03e-06}}
\newcommand{\eventfulANTOkwp}{\num{0.671}}
\newcommand{\eventfulORTHkwp}{\num{3.37e-05}}
\newcommand{\eventfulNCONkwp}{\num{1.33e-08}}
\newcommand{\eventfulIBALkwp}{\num{2.61e-06}}
\newcommand{\uneventfulAPPRkwp}{\num{0.0141}}
\newcommand{\uneventfulUNDRkwp}{\num{0.000967}}
\newcommand{\uneventfulCLARkwp}{\num{0.00332}}
\newcommand{\uneventfulANTOkwp}{\num{0.0268}}
\newcommand{\uneventfulORTHkwp}{\num{2.25e-09}}
\newcommand{\uneventfulNCONkwp}{\num{0.0178}}
\newcommand{\uneventfulIBALkwp}{\num{4.18e-08}}
\newcommand{\calmAPPRkwp}{\num{3.2e-11}}
\newcommand{\calmUNDRkwp}{\num{0.000213}}
\newcommand{\calmCLARkwp}{\num{0.414}}
\newcommand{\calmCONNkwp}{\num{0.133}}
\newcommand{\calmIBALkwp}{\num{0.00204}}
\newcommand{\chaoticAPPRkwp}{\num{2.53e-06}}
\newcommand{\chaoticUNDRkwp}{\num{2.19e-07}}
\newcommand{\chaoticCLARkwp}{\num{0.0303}}
\newcommand{\chaoticCONNkwp}{\num{0.126}}
\newcommand{\chaoticIBALkwp}{\num{0.771}}
\newcommand{\vibrantAPPRkwp}{\num{2.77e-11}}
\newcommand{\vibrantUNDRkwp}{\num{2.47e-07}}
\newcommand{\vibrantCLARkwp}{\num{7.61e-09}}
\newcommand{\vibrantCONNkwp}{\num{2.21e-08}}
\newcommand{\vibrantIBALkwp}{\num{0.0167}}
\newcommand{\monotonousAPPRkwp}{\num{0.425}}
\newcommand{\monotonousUNDRkwp}{\num{1.33e-08}}
\newcommand{\monotonousCLARkwp}{\num{0.269}}
\newcommand{\monotonousCONNkwp}{\num{0.608}}
\newcommand{\monotonousIBALkwp}{\num{0.292}}
\newcommand{\pleasantUNDRnaafangGTruenhuu}{${p<\num{0.001}}$}
\newcommand{\pleasantUNDRnaafangGTsanorhuu}{${p<\num{0.001}}$}
\newcommand{\pleasantUNDRruenhuuGTsanorhuu}{${p=\num{0.0494}}$}
\newcommand{\annoyingAPPRnaaramkaanGTrakaaihuu}{${p<\num{0.001}}$}
\newcommand{\annoyingUNDRnaaramkaanGTrakaaihuu}{${p<\num{0.001}}$}
\newcommand{\eventfulAPPRwunwaaiGTmiiarai}{${p=\num{0.202}}$}
\newcommand{\eventfulAPPRuekgatuekGTmiiarai}{${p=\num{0.0274}}$}
\newcommand{\eventfulAPPRuekgatuekGTwunwaai}{${p\approx\num{1}}$}
\newcommand{\eventfulUNDRwunwaaiGTmiiarai}{${p=\num{0.00204}}$}
\newcommand{\eventfulUNDRmiiaraiGTuekgatuek}{${p=\num{0.0468}}$}
\newcommand{\eventfulUNDRwunwaaiGTuekgatuek}{${p<\num{0.001}}$}
\newcommand{\eventfulCLARmiiaraiGTwunwaai}{${p<\num{0.001}}$}
\newcommand{\eventfulCLARmiiaraiGTuekgatuek}{${p<\num{0.001}}$}
\newcommand{\eventfulCLARuekgatuekGTwunwaai}{${p=\num{0.284}}$}
\newcommand{\eventfulORTHmiiaraiGTwunwaai}{${p<\num{0.001}}$}
\newcommand{\eventfulORTHmiiaraiGTuekgatuek}{${p=\num{0.0552}}$}
\newcommand{\eventfulORTHuekgatuekGTwunwaai}{${p=\num{0.0255}}$}
\newcommand{\eventfulNCONmiiaraiGTwunwaai}{${p<\num{0.001}}$}
\newcommand{\eventfulNCONmiiaraiGTuekgatuek}{${p<\num{0.001}}$}
\newcommand{\eventfulNCONuekgatuekGTwunwaai}{${p=\num{0.351}}$}
\newcommand{\eventfulIBALmiiaraiGTwunwaai}{${p<\num{0.001}}$}
\newcommand{\eventfulIBALmiiaraiGTuekgatuek}{${p=\num{0.0775}}$}
\newcommand{\eventfulIBALuekgatuekGTwunwaai}{${p=\num{0.0016}}$}
\newcommand{\uneventfulAPPRsangopGTriiap}{${p=\num{0.261}}$}
\newcommand{\uneventfulAPPRmaimiiaraiGTsangop}{${p=\num{0.599}}$}
\newcommand{\uneventfulAPPRmaimiiaraiGTriiap}{${p=\num{0.00981}}$}
\newcommand{\uneventfulUNDRsangopGTriiap}{${p<\num{0.001}}$}
\newcommand{\uneventfulUNDRsangopGTmaimiiarai}{${p=\num{0.437}}$}
\newcommand{\uneventfulUNDRmaimiiaraiGTriiap}{${p=\num{0.0441}}$}
\newcommand{\uneventfulCLARriiapGTsangop}{${p\approx\num{1}}$}
\newcommand{\uneventfulCLARmaimiiaraiGTsangop}{${p=\num{0.00279}}$}
\newcommand{\uneventfulCLARmaimiiaraiGTriiap}{${p=\num{0.0509}}$}
\newcommand{\uneventfulANTOsangopGTriiap}{${p=\num{0.0477}}$}
\newcommand{\uneventfulANTOsangopGTmaimiiarai}{${p\approx\num{1}}$}
\newcommand{\uneventfulANTOmaimiiaraiGTriiap}{${p=\num{0.0645}}$}
\newcommand{\uneventfulORTHriiapGTsangop}{${p<\num{0.001}}$}
\newcommand{\uneventfulORTHmaimiiaraiGTsangop}{${p<\num{0.001}}$}
\newcommand{\uneventfulORTHmaimiiaraiGTriiap}{${p=\num{0.0221}}$}
\newcommand{\uneventfulNCONriiapGTsangop}{${p\approx\num{1}}$}
\newcommand{\uneventfulNCONmaimiiaraiGTsangop}{${p=\num{0.017}}$}
\newcommand{\uneventfulNCONmaimiiaraiGTriiap}{${p=\num{0.138}}$}
\newcommand{\uneventfulIBALriiapGTsangop}{${p<\num{0.001}}$}
\newcommand{\uneventfulIBALmaimiiaraiGTsangop}{${p<\num{0.001}}$}
\newcommand{\uneventfulIBALriiapGTmaimiiarai}{${p\approx\num{1}}$}
\newcommand{\calmAPPRsangopGTpornklaai}{${p<\num{0.001}}$}
\newcommand{\calmAPPRpornklaaiGTsabaaihuu}{${p<\num{0.001}}$}
\newcommand{\calmAPPRsangopGTsabaaihuu}{${p<\num{0.001}}$}
\newcommand{\calmUNDRsangopGTpornklaai}{${p=\num{0.279}}$}
\newcommand{\calmUNDRpornklaaiGTsabaaihuu}{${p=\num{0.0209}}$}
\newcommand{\calmUNDRsangopGTsabaaihuu}{${p<\num{0.001}}$}
\newcommand{\calmIBALsangopGTpornklaai}{${p=\num{0.485}}$}
\newcommand{\calmIBALpornklaaiGTsabaaihuu}{${p=\num{0.0718}}$}
\newcommand{\calmIBALsangopGTsabaaihuu}{${p=\num{0.00108}}$}
\newcommand{\chaoticAPPRbpanbpuuanGTjorjae}{${p=\num{0.00157}}$}
\newcommand{\chaoticAPPRyungyerngGTjorjae}{${p<\num{0.001}}$}
\newcommand{\chaoticAPPRwunwaaiGTjorjae}{${p<\num{0.001}}$}
\newcommand{\chaoticAPPRyungyerngGTbpanbpuuan}{${p\approx\num{1}}$}
\newcommand{\chaoticAPPRwunwaaiGTbpanbpuuan}{${p=\num{0.229}}$}
\newcommand{\chaoticAPPRwunwaaiGTyungyerng}{${p=\num{0.556}}$}
\newcommand{\chaoticUNDRbpanbpuuanGTjorjae}{${p=\num{0.0542}}$}
\newcommand{\chaoticUNDRyungyerngGTjorjae}{${p=\num{0.0026}}$}
\newcommand{\chaoticUNDRwunwaaiGTjorjae}{${p<\num{0.001}}$}
\newcommand{\chaoticUNDRyungyerngGTbpanbpuuan}{${p\approx\num{1}}$}
\newcommand{\chaoticUNDRwunwaaiGTbpanbpuuan}{${p<\num{0.001}}$}
\newcommand{\chaoticUNDRwunwaaiGTyungyerng}{${p=\num{0.017}}$}
\newcommand{\chaoticCLARbpanbpuuanGTjorjae}{${p\approx\num{1}}$}
\newcommand{\chaoticCLARyungyerngGTjorjae}{${p\approx\num{1}}$}
\newcommand{\chaoticCLARjorjaeGTwunwaai}{${p=\num{0.197}}$}
\newcommand{\chaoticCLARyungyerngGTbpanbpuuan}{${p\approx\num{1}}$}
\newcommand{\chaoticCLARbpanbpuuanGTwunwaai}{${p=\num{0.133}}$}
\newcommand{\chaoticCLARyungyerngGTwunwaai}{${p=\num{0.035}}$}
\newcommand{\vibrantAPPRmiichiiwitchiiwaaGTkuekkak}{${p=\num{0.401}}$}
\newcommand{\vibrantAPPRkuekkakGTsotsai}{${p=\num{0.383}}$}
\newcommand{\vibrantAPPRkuekkakGTwuewaa}{${p<\num{0.001}}$}
\newcommand{\vibrantAPPRmiichiiwitchiiwaaGTsotsai}{${p=\num{0.00182}}$}
\newcommand{\vibrantAPPRmiichiiwitchiiwaaGTwuewaa}{${p<\num{0.001}}$}
\newcommand{\vibrantAPPRsotsaiGTwuewaa}{${p<\num{0.001}}$}
\newcommand{\vibrantUNDRmiichiiwitchiiwaaGTkuekkak}{${p\approx\num{1}}$}
\newcommand{\vibrantUNDRsotsaiGTkuekkak}{${p\approx\num{1}}$}
\newcommand{\vibrantUNDRkuekkakGTwuewaa}{${p<\num{0.001}}$}
\newcommand{\vibrantUNDRsotsaiGTmiichiiwitchiiwaa}{${p\approx\num{1}}$}
\newcommand{\vibrantUNDRmiichiiwitchiiwaaGTwuewaa}{${p<\num{0.001}}$}
\newcommand{\vibrantUNDRsotsaiGTwuewaa}{${p<\num{0.001}}$}
\newcommand{\vibrantCLARkuekkakGTmiichiiwitchiiwaa}{${p\approx\num{1}}$}
\newcommand{\vibrantCLARsotsaiGTkuekkak}{${p=\num{0.426}}$}
\newcommand{\vibrantCLARwuewaaGTkuekkak}{${p<\num{0.001}}$}
\newcommand{\vibrantCLARsotsaiGTmiichiiwitchiiwaa}{${p=\num{0.224}}$}
\newcommand{\vibrantCLARwuewaaGTmiichiiwitchiiwaa}{${p<\num{0.001}}$}
\newcommand{\vibrantCLARwuewaaGTsotsai}{${p<\num{0.001}}$}
\newcommand{\vibrantCONNmiichiiwitchiiwaaGTkuekkak}{${p\approx\num{1}}$}
\newcommand{\vibrantCONNkuekkakGTsotsai}{${p=\num{0.12}}$}
\newcommand{\vibrantCONNkuekkakGTwuewaa}{${p<\num{0.001}}$}
\newcommand{\vibrantCONNmiichiiwitchiiwaaGTsotsai}{${p=\num{0.0795}}$}
\newcommand{\vibrantCONNmiichiiwitchiiwaaGTwuewaa}{${p<\num{0.001}}$}
\newcommand{\vibrantCONNsotsaiGTwuewaa}{${p=\num{0.00524}}$}
\newcommand{\vibrantIBALmiichiiwitchiiwaaGTkuekkak}{${p=\num{0.0286}}$}
\newcommand{\vibrantIBALsotsaiGTkuekkak}{${p\approx\num{1}}$}
\newcommand{\vibrantIBALwuewaaGTkuekkak}{${p\approx\num{1}}$}
\newcommand{\vibrantIBALmiichiiwitchiiwaaGTsotsai}{${p=\num{0.0779}}$}
\newcommand{\vibrantIBALmiichiiwitchiiwaaGTwuewaa}{${p=\num{0.0569}}$}
\newcommand{\vibrantIBALsotsaiGTwuewaa}{${p\approx\num{1}}$}
\newcommand{\monotonousUNDRnaabueaGTjuedjued}{${p<\num{0.001}}$}
\newcommand{\monotonousUNDRjuedjuedGTnuuainuuai}{${p=\num{0.0424}}$}
\newcommand{\monotonousUNDRjuedjuedGTuuaiuuai}{${p=\num{0.944}}$}
\newcommand{\monotonousUNDRnaabueaGTnuuainuuai}{${p<\num{0.001}}$}
\newcommand{\monotonousUNDRnaabueaGTuuaiuuai}{${p<\num{0.001}}$}
\newcommand{\monotonousUNDRuuaiuuaiGTnuuainuuai}{${p\approx\num{1}}$}

\iflineno\linenumbers\fi

\newcommand*{\papertitle}{Quantitative Evaluation Approach for Translation of Perceptual Soundscape Attributes: Initial Application to the Thai Language}
\shorttitle{\papertitle}    
\shortauthors{K. N. Watcharasupat et al.}  

\title[mode=title]{\papertitle}  
\author[eee]{Karn N. Watcharasupat}[orcid=0000-0002-3878-5048]
\ead{karn001@e.ntu.edu.sg}
\corref{c}\cortext[c]{Corresponding author}
\credit{Conceptualization, Methodology, Formal analysis, Investigation, Writing - Original Draft, Writing - Review \& Editing, Resources, Project administration}

\author[soh]{Sureenate Jaratjarungkiat}[orcid=0000-0001-8295-6775]
\ead{sureenate@ntu.edu.sg}
\credit{Methodology, Investigation, Writing - Review \& Editing, Resources, Project administration}

\author[eee]{Bhan Lam}[orcid=0000-0001-5193-6560]
\ead{bhanlam@ntu.edu.sg}
\credit{Conceptualization, Methodology, Formal analysis, Writing - Review \& Editing, Project administration}

\author[chula]{Sujinat Jitwiriyanont}[orcid=0000-0002-5785-9882]
\ead{sujinat.j@chula.ac.th}
\credit{Investigation, Writing - Review \& Editing}

\author[eee]{Kenneth Ooi}[orcid=0000-0001-5629-6275]
\ead{wooi002@e.ntu.edu.sg}
\credit{Methodology, Formal analysis, Writing - Review \& Editing}

\author[eee]{Zhen-Ting Ong}[orcid=0000-0002-1249-4760]
\ead{ztong@ntu.edu.sg}
\credit{Resources, Project administration}

\author[tu]{Nitipong Pichetpan}[orcid=0000-0001-8934-2039]
\ead{pnitipon@tu.ac.th}
\credit{Investigation, Writing - Review \& Editing}

\author[ku]{Kanyanut Akaratham}
\ead{kanyanut.ak@live.ku.th}
\credit{Resources, Writing - Review \& Editing}

\author[cls]{Titima Suthiwan}[orcid=0000-0002-2273-8239]
\ead{clsts@nus.edu.sg}
\credit{Investigation}

\author[tu]{Monthita Rojtinnakorn}[orcid=0000-0002-5785-9882]
\ead{monthita@tu.ac.th}
\credit{Investigation}

\author[eee]{Woon-Seng Gan}[orcid=0000-0002-7143-1823]
\ead{ewsgan@ntu.edu.sg}
\credit{Writing - Review \& Editing, Supervision, Funding acquisition}

\affiliation[eee]{
    organization={%%
        % Digital Signal Processing Laboratory,
        School of Electrical and Electronic Engineering, 
        Nanyang Technological University%
    },
    addressline={50 Nanyang Ave, S2-B4a-03}, 
    % city={Singapore},
    postcode={639798}, 
    % state={Singapore},
    country={Singapore}
}

\affiliation[soh]{
    organization={%
        Centre for Modern Languages,
        School of Humanities,   
        Nanyang Technological University%
    },
    addressline={48 Nanyang Ave}, 
    % city={Singapore},
    postcode={639818}, 
    % state={Singapore},
    country={Singapore}
}

\affiliation[chula]{
    organization={%
        Department of Linguistics and Southeast Asian Linguistics Research Unit,   
        Faculty of Arts, 
        Chulalongkorn University%
    },
    addressline={%
        254 Phayathai Rd, Wang Mai, 
        Pathum Wan District%
    },
    city={Bangkok},
    postcode={10330}, 
    country={Thailand}
}

\affiliation[tu]{
    organization={%
        Faculty of Liberal Arts, 
        Thammasat University%
    },
    addressline={%
        99 Moo 18 Phahonyothin Rd, Khlong Nueng,  
        Khlong Luang District%
    }, 
    city={Pathum Thani},
    postcode={12121}, 
    country={Thailand}
}

\affiliation[ku]{
    organization={%
        Department of Psychology,
        Faculty of Social Sciences,
        Kasetsart University%
    },
    addressline={%
        50 Ngamwongwan Rd, Lat Yao, Chatuchak District% 
    },
    city={Bangkok},
    postcode={10900},
    country={Thailand}
}

\affiliation[cls]{
    organization={%
        Centre for Language Studies,   
        Faculty of Arts and Social Sciences,
        National University of Singapore%
    },
    addressline={9 Arts Link},
    % city={Singapore},
    postcode={117572}, 
    % state={Singapore},
    country={Singapore}
}

%title footnote
\tnotemark[1] 
\tnotetext[1]{The research protocols used in this research were approved by the institutional review board of Nanyang Technological University (NTU), Singapore [IRB-2021-293].}

\begin{abstract}
Translation of perceptual soundscape attributes from one language to another remains a challenging task that requires a high degree of fidelity in both psychoacoustic and psycholinguistic senses across the target population. Due to the inherently subjective nature of human perception, translating soundscape attributes using only small focus group discussion\added{s} or expert panels could lead to translations with psycholinguistic meanings that, in a non-expert setting, deviate or distort from that of the source language. In this work, we present a quantitative evaluation method based on the circumplex model of soundscape perception to assess the overall translation quality. \deleted{across a set of criteria,} \added{By establishing a set of criteria for evaluating the linguistic and psychometric properties of the translation candidates,
% --- namely, \textit{appropriateness}, \textit{understandability}, \textit{clarity}, \textit{orthogonal unbiasedness}, \textit{antipodal antonymity}, \textit{connotativeness}, \textit{nonconnotativeness}, and \textit{implicative balance} --- 
statistical analyses can be performed to objectively assess specific strengths and weaknesses of the translation candidates before committing to listening tests or more involved validation experiments.}
As an initial application domain, we demonstrated the use of the quantitative evaluation framework in the context of an English-to-Thai translation of soundscape attributes. \added{A total of 31 participants who are bilingual in English and Thai were recruited to assess the translation candidates. Subsequent statistical analysis of the evaluation scores revealed acoustico-psycholinguistic properties of the translation candidates which were not previously identified by the expert panel and facilitated a more objective selection of the final translations for subsequent usage. Additionally, with specific biases of the final translations determined numerically, mathematical and statistical techniques for corrections of the survey data may be employed in the future to improve cross-lingual compatibility in soundscape evaluation.} 
\end{abstract}

% \begin{highlights}
% \item 
% \item 
% \item 
% \end{highlights}

% Keywords

\maketitle
% \setkeys{Gin}{draft=true}
% Main text
\section{Introduction}\label{sec:intro}

The standardization of soundscape protocols in the ISO 12913 series of standards \citep{ISO2014ISOFramework, ISO2018ISO/TSRequirements, ISO2019ISO/TSAnalysis} has greatly unified the reporting standards and data collection methodology in soundscape studies. However, the protocols for data collection were only standardized in English, leaving the use of soundscape terminologies largely unstandardized in regions of the world where English is not the main language. Since soundscape studies, by definition, are concerned with the relationships between an acoustic environment and the human experience \textit{in context} \citep{ISO2014ISOFramework}, soundscape studies are highly intertwined with the socio-cultural context not only of the acoustic environment itself, but also of the language medium in which the studies are conducted. The lack of standardization in soundscape has posed a serious challenge in the transferability and comparability of soundscape knowledge along the linguistic borders.

% \subsection{Related Work}
\deleted[comment={moved to~\S2}]{
Prior to the publication of ISO/TS 12913-2:2018 and 12913-3:2019, much of the work on soundscape assessment has been reliant on the Swedish Soundscape-Quality Protocol (SSQP) by \mbox{\citet{Axelsson2009AQuality, Axelsson2010APerception, Axelsson2012TheProtocol}}, which, together with the work of \mbox{\citet{Cain2013TheSoundscape}}, would later become the basis for the two-dimensional model in the ISO/TS 12913-3:2019. The SSQP itself has previously been translated into 10 to 15 languages, albeit without experimental validation to ensure interlingual compatibility. Unsurprisingly, a cross-national study by \mbox{\citet{Jeon2018AExperiments}} found statistically significant differences in soundscape assessment across France, Korea, and Sweden, highlighting important comparability issues of the soundscape assessment scales across languages. These linguistic issues are also present in the findings of studies conducted in Japan \mbox{\citep{Nagahata2018LinguisticResearch, Nagahata2019ExaminationJapanese}}, and a French-speaking region of Canada \mbox{\citep{Tarlao2016ComparingMontreal,Tarlao2021InvestigatingModeling}}.
}

After the publication of the ISO\added{/TS} 12913-2:2018 and 12913-3:2019 \added{standards}, an international collaboration was initiated amongst soundscape researchers in order to develop experimentally validated and cross-linguistically compatible sets of translations of the soundscape attributes. The collaboration, named \textit{Soundscape Attribute Translation Project} (SATP), is currently working on 19 languages, with several whose experimental validations have been completed. Generally, most working groups followed a two-stage structure where the first is concerned with developing a set of provisional translations, while the second is concerned with validating the translations via a listening test \citep{Aletta2020SoundscapeLanguages}. 

Due to the lack of `gold standard' procedures for translation in the field of soundscapes, provisional translation stage for Mandarin Chinese, Yue Chinese, Croatian, Dutch, French, Italian, Korean, Spanish, Swedish, Turkish, Vietnamese \citep{Aletta2020SoundscapeLanguages}, and Bahasa Indonesia \citep{Sudarsono2022TheStudy}, have largely relied on expert panels and/or focus group discussions, and, if any, previously soundscape works involving translations in their respective languages. However, solely relying on an expert panel for the translation process may not always produce a set of translations that is compatible with the English standards. As seen in the Indonesian study \citep{Sudarsono2022TheStudy}, significant deviations from the English results were found for the translations of \textit{eventful}, \textit{uneventful}, and \textit{chaotic} during the experimental validation. \added{Recently, \citet{Antunes2021ValidatedAssessment} took a questionnaire approach for the binational translation process for Portuguese, drawing participants from both Portugal and Brazil, to more objectively assess the translation quality. However, their approach only evaluated the overall suitability of the translation candidates, but not specific psychometric properties.}

Admittedly, it is possible to approach the translation process and validation process in an iterative manner. However, this is usually neither desirable nor practical, as conducting listening tests is a time-consuming and labor-intensive process. Moreover, testing multiple sets of candidate translations in one sitting is also often not advisable, as the number of candidate translations is limited by experimental fatigue of the participants \citep{Schatz2012TheRatings, Schwarz2016EffectsTests}. 

\deleted[comment={moved to~\S2}]{Interestingly, the Portuguese provisional translation process \mbox{\citep{Antunes2021ValidatedAssessment}} took a more structured and slightly more quantitative approach with Stage 1 divided into two substages. Although the first substage concerning the initial selection process of candidate translations remains inevitably dependent on an expert panel, in the second substage, an online questionnaire administered cross-nationally in Portugal and Brazil was employed to finalize their set of provisional translations before proceeding to the listening test stage. For each soundscape attribute, the questionnaire item starts with a sentence describing an acoustic environment with a perceptual quality of the attribute, followed by asking which of the candidate translations is considered most ``suitable''. It is important to note that the participants are also given a free response box for each item to suggest another translation. The results of the questionnaire provided the Portuguese working group with quantitative insights on the suitability of each candidate translation, as well as cross-cultural differences in the use of the Portuguese language between Portugal and Brazil.}

\subsection{Contributions}

In this work, we build on the insights from the Portuguese translation process in \citet{Antunes2021ValidatedAssessment} and propose an extended quantitative evaluation framework for the translation of soundscape attributes. Crucially, instead of a choice-based evaluation of overall suitability, \deleted{used in \mbox{\citet{Antunes2021ValidatedAssessment}},} our framework utilizes a score-based approach to assess the quality of a translation across multiple criteria concerning the appropriateness of translation, understandability, clarity, and its linguistic relationships to other soundscape attributes. In particular, the use of multiple criteria in a score-based framework allows for the use of statistical analysis in the selection process, as well as the identification of particular strengths or weaknesses each candidate translation may possess. Overall, the framework seeks to identify a set of provisional translations which preserves not only the original meanings, but also the inter-attribute relationship between the eight soundscape descriptors, \added[comment=R2.1]{in a manner that aims to minimize linguistic discrepancies but preserves cultural differences.}

The quantitative framework proposed in this paper has been used for the translation process for the Thai language (ISO 639-3: \textsc{tha}) and Bahasa Melayu (ISO 639-3: \textsc{zsm}). In this work, we will focus on the design of the quantitative framework, and its application to the Thai language. Due to the cross-national nature of Bahasa Melayu, a full discussion of its translation process is beyond the scope of this paper and will be addressed in detail in a separate work.

\subsection{Terminology}

\begin{figure}[t]
    \centering
    \includegraphics[width=0.9\columnwidth]{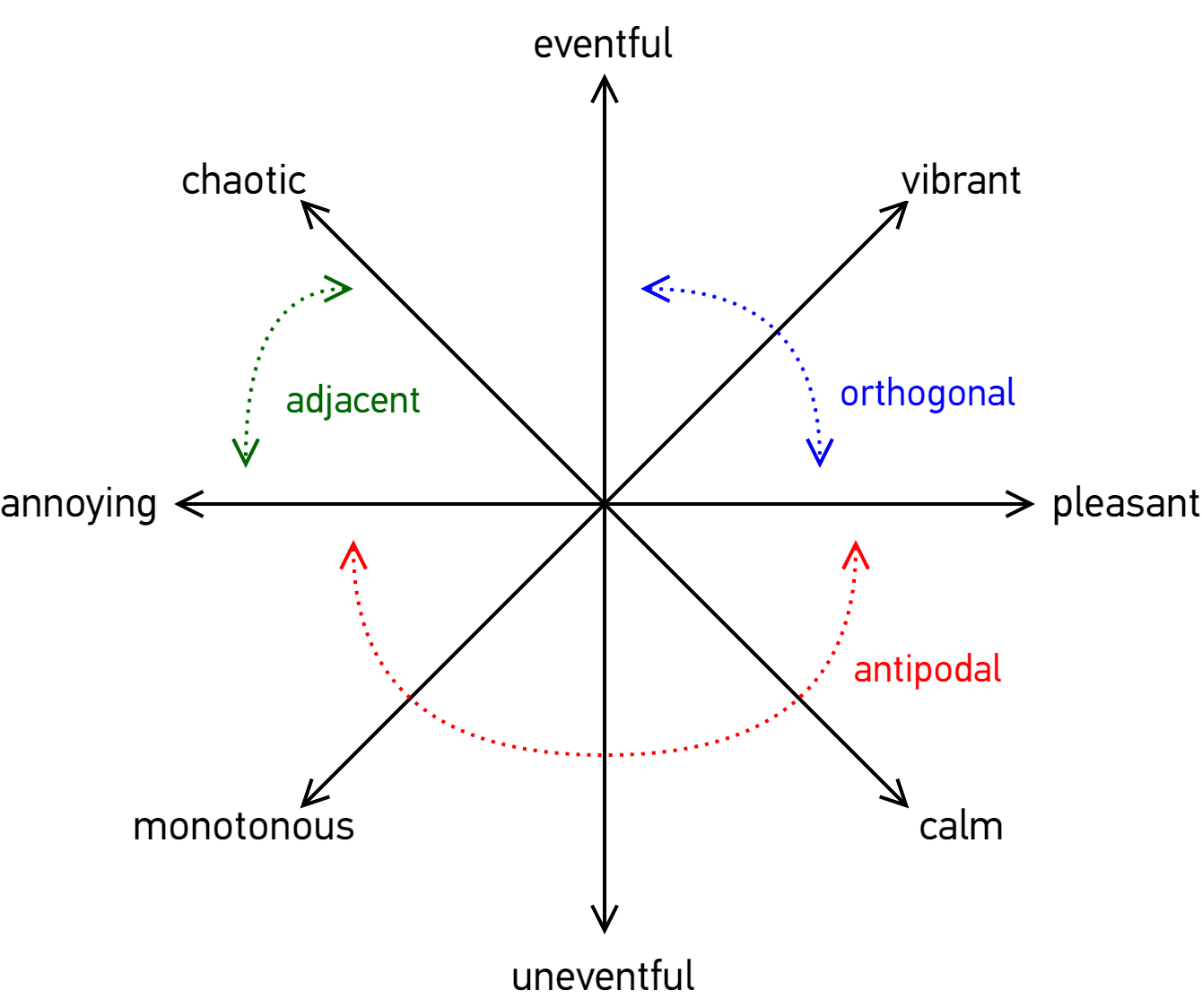}
    \caption{The circumplex model of soundscape attributes, as per ISO/TS 12913-3:2019 \citep{ISO2019ISO/TSAnalysis}. The curved arrows illustrate the terms used to describe relations between soundscape attributes in this paper.}
    \label{fig:circ}
\end{figure}

This paper largely follows the soundscape attribute terminology used in the ISO/TS 12913 standards \citep{ISO2014ISOFramework, ISO2018ISO/TSRequirements, ISO2019ISO/TSAnalysis}. Additionally, we introduce terms to describe the relationship between soundscape attributes in the \added{theoretical} circumplex model as follows. Attributes located \added{theoretically} \SI{45}{\degree} from each other are considered \textit{adjacent}. Attributes located \added{theoretically} \SI{90}{\degree} from each other are considered \textit{orthogonal}, and attributes located \added{theoretically} \SI{180}{\degree} from each other are considered \textit{antipodal}. Additionally, \textit{clockwise} and \textit{counterclockwise} qualifiers are used to specify the relative location of an attribute with respect to another based on their locations on the circumplex model. \Cref{fig:circ} illustrates the circumplex model of soundscape attributes and the relation\added{al} terminology used in this paper.

\subsection{Organization}

This paper is organized as follows. \added{\Cref{sec:discussion} discusses related work.} \Cref{sec:vq} discusses the design of the proposed quantitative evaluation framework.
\added{Application of the proposed framework to the Thai language is detailed in} \Cref{sec:init} and \Cref{sec:quant},\deleted{discuss the application of the proposed framework into the Thai language} with the former discussing the initial expert translation phase, and the latter discussing the validation questionnaire. 
Finally, \Cref{sec:conclusion} concludes the paper and lays out future work. 

For readability, Thai words used in this paper are followed by \deleted{its}\added{their} International Phonetic Alphabet (IPA) transcription \added{\citep[adapted from][]{Tingsabadh1993Thai,Nacaskul2013TheThai}}, and, where appropriate, also by \deleted{its}\added{their} literal meaning following the \textit{lit.} indicator.
\begin{sadded}

\section{Related Work}\label{sec:discussion}

\comment{moved from \S1}

\begin{smoved}
Prior to the publication of ISO/TS 12913-2:2018 and 12913-3:2019, much of the work on soundscape assessment has been reliant on the Swedish Soundscape-Quality Protocol (SSQP) by \citet{Axelsson2009AQuality, Axelsson2010APerception, Axelsson2012TheProtocol}, which, together with the work of \citet{Cain2013TheSoundscape}, would later become the basis for the two-dimensional model in the ISO/TS 12913-3:2019. The SSQP itself has previously been translated into 10 to 15 languages, albeit without experimental validation to ensure interlingual compatibility. Unsurprisingly, a cross-national study by \citet{Jeon2018AExperiments} found statistically significant differences in soundscape assessment across France, Korea, and Sweden, highlighting important comparability issues of the soundscape assessment scales across languages. These linguistic issues are also present in the findings of studies conducted in Japan \citep{Nagahata2018LinguisticResearch, Nagahata2019ExaminationJapanese}, and \added{Montreal,} a French-speaking region of Canada \citep{Tarlao2016ComparingMontreal,Tarlao2021InvestigatingModeling}.
\end{smoved}

\subsection{Cross-lingual soundscape assessment}

However, it is also well known that some differences \added{in} soundscape assessment across populations can be attributed to cultural differences. In fact, distinguishing the sources of measured differences in soundscape and noise evaluation across different populations has long been an open problem in the field of psychometry \citep{Sperber1994Cross-culturalValidation} and subsequently soundscapes, even prior to the introduction of the SSQP. Although a number of cross-cultural, cross-national, and cross-lingual studies related to the perception of sound have been conducted, the effects of translations and cultures have yet to be extensively discussed in a systematic manner \citep{Kuwano1999ANoise,Phan2008AnnoyanceJapanese,Hansen2009SemanticComparison,Yu2014SoundscapeTaiwan,Jeon2018AExperiments,Deng2020Cross-NationalCroatia,Mohamed2020IndoorTurkey,Tarlao2016ComparingMontreal}. To the best of our knowledge, a recent work by \citet{Tarlao2021InvestigatingModeling} is the first and only attempt so far at studying the statistical construct of soundscape assessment tools across translations.

In most studies, the studied population `groups' are both geographically and linguistically distinct. 
% A factor analysis performed in \citet{Kuwano1999ANoise} revealed both semantic similarities and differences in environmental noise descriptors in five cities across Japan (in Japanese), the United States (in English), China (in Chinese), and Germany (in German for Oldenburg, and English for Munich). 
A study by \citet{Phan2008AnnoyanceJapanese} between Japan (in Japanese) and Vietnam (in Vietnamese) found differences in the perception of traffic noise between the two populations. \citeauthor{Phan2008AnnoyanceJapanese} noted that some differences could be due to differences in the intensity of linguistic modifiers between Vietnamese and Japanese. A cross-national study in France, Korea, and Sweden by \citet{Jeon2018AExperiments} using translations of the SSQP into French, Korean, and Swedish found some evidence of socio-cultural differences in soundscape perception, although the authors remarked that linguistic factors likely also contributed to the differences in perceptual responses. 
\citet{Yu2014SoundscapeTaiwan} conducted a cross-national using questionnaire surveys in the United Kingdom (in English) and Taiwan (in Chinese), while \citet{Deng2020Cross-NationalCroatia} did so in China (in Chinese) and Croatia (in Croatian and English). Both \citet{Yu2014SoundscapeTaiwan} and \citet{Deng2020Cross-NationalCroatia} and differences in soundscape perception between their respective study sites. However, the two studies did not discuss the extent to which linguistic differences contributed to the differences in survey responses.

On the other hand, in \citet{Tarlao2016ComparingMontreal,Tarlao2021InvestigatingModeling} and \citet{Mohamed2020IndoorTurkey}, the population `groups' share the same geographical location but are ethnically and/or linguistically distinct. \citet{Mohamed2020IndoorTurkey} studied the difference in indoor soundscape perceptions between the Arab and Turkish ethnic groups within Turkey, using questionnaire surveys in Arabic and Turkish, respectively. The Turkey study found a number of differences in soundscape perception, which the authors attributed to cultural differences. However, despite the survey being carried out in two languages, the effects of the translations were not discussed in the paper. In another work, \citet{Tarlao2016ComparingMontreal,Tarlao2021InvestigatingModeling} studied the difference between the English and French versions of the SSQP within the French-speaking city of Montreal in Canada. It could thus be argued that linguistic effects may be more salient in studies where the population groups are culturally closer to one another, as such when there are geographically colocated.

Crucially, \citet{Tarlao2021InvestigatingModeling} demonstrated that the SSQP exhibits metric invariance (i.e., that the underlying \textit{pleasantness} and \textit{eventfulness} latent factors of soundscapes are shared between the French and English groups) but not scalar invariance (i.e., that each soundscape attribute scale cannot be treated as equivalent between the English and French versions). Unfortunately, no similar study exists for the ISO/TS 12913 version of soundscape evaluation yet, as \citeauthor{Tarlao2021InvestigatingModeling} relied on past data collected prior to the introduction of ISO/TS 12913-2:2018 and 12913-3:2019.

\subsection{Distinguishing the cultural and linguistic effects}

% \setkeys{Gin}{draft=false}
\begin{figure*}[t]
    \centering
    \includegraphics[width=\textwidth]{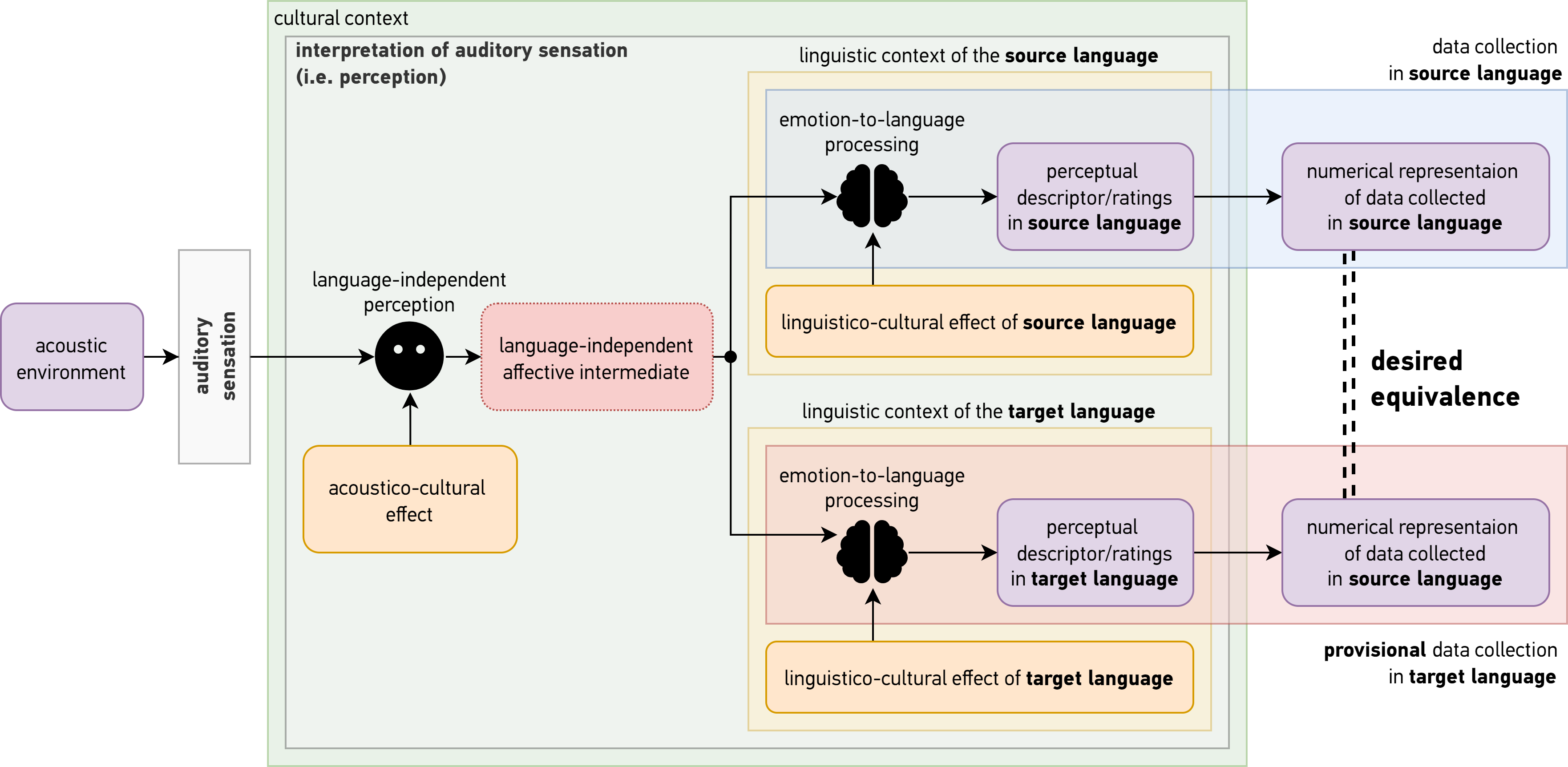}
    \caption{A multilingual model of sound perception when situated in a hypothetical unicultural context. Extended from the ISO~12913-1:2014 perceptual construct of soundscape \citep[Fig.\ 1]{ISO2014ISOFramework}.}
    \label{fig:perception}
\end{figure*}

Due to cross-lingual incompatibility experienced with translations of the SSQP, the SATP initiative was formed following the publication of ISO/TS 12913-2:2018 and 12913-3:2019 to ensure translations of the new standards are more comparable across languages. In other words, the goal of the SATP initiative \citep{Aletta2020SoundscapeLanguages} could be considered as an endeavour towards multilingual scalar invariance of the soundscape attributes. 

However, achieving scalar invariance in the most naive sense would perhaps defeat the purpose of the initiative, as the instrument would also erase the cultural differences in perception that is of significant importance in soundscape research, by the virtue of culture being a part of the \textit{context} as defined in ISO 12913:2014 \citep{ISO2014ISOFramework}. In a hypothetical unicultural but multilingual world, naive scalar invariance across translations is indeed desirable. However, the real world is both multilingual and multicultural; achieving naive scalar invariance by the mean of validated translations may mean that there was an overcalibration of the psychometric scale such that not only were linguistic differences desirably erased, but cultural differences were also undesirably erased. A very difficult open question is thus how to remove \textit{only} linguistic differences from the translations when languages are very much established to be intertwined with their cultures \citep{Harzing2002TheCountries}. 

In \citet{Juslin2016PrevalenceCultures} and \citet{Huang2018AnMusic}, evidence of cultural modulation effects on physiological responses to acoustic stimuli has been demonstrated. Since direct measurements of physiological responses would bypass the language-processing mechanism needed in survey-based studies, this demonstrated some evidence of language-independent acoustico-cultural effects. It has also been established that language serves as a medium of communication of emotional \textit{concept}, forming a specialized pathway in the brain whereby the linguistic medium itself has a constructive effect on the eventual perception thus the measured responses \citep{Kotz2011EmotionBrain, Lindquist2015TheConstructionism}. The latter pathway itself is also modulated by the cultural context in which the lexicosemantics of the linguistic medium is situated \citep{Goddard2015WordsMeaning}. As a result, it may be impossible to remove linguistic effects without any distortion of cultural effects. However, it may be possible to remove only the linguistico-cultural effects with minimal distortion on the acoustico-cultural effects.

In a simplified model, we can consider the process of interpreting auditory sensation (i.e., perception) as a two-step pathway. The first step involves a culture-dependent but language-independent perception, resulting in an affective intermediate. 
To communicate the affective intermediate to the data collection medium (e.g., survey), the emotion-to-language pathway is involved, upon which the linguistico-cultural modulation effect is triggered.

\begin{figure*}[t]
    \centering
    \includegraphics[width=\textwidth]{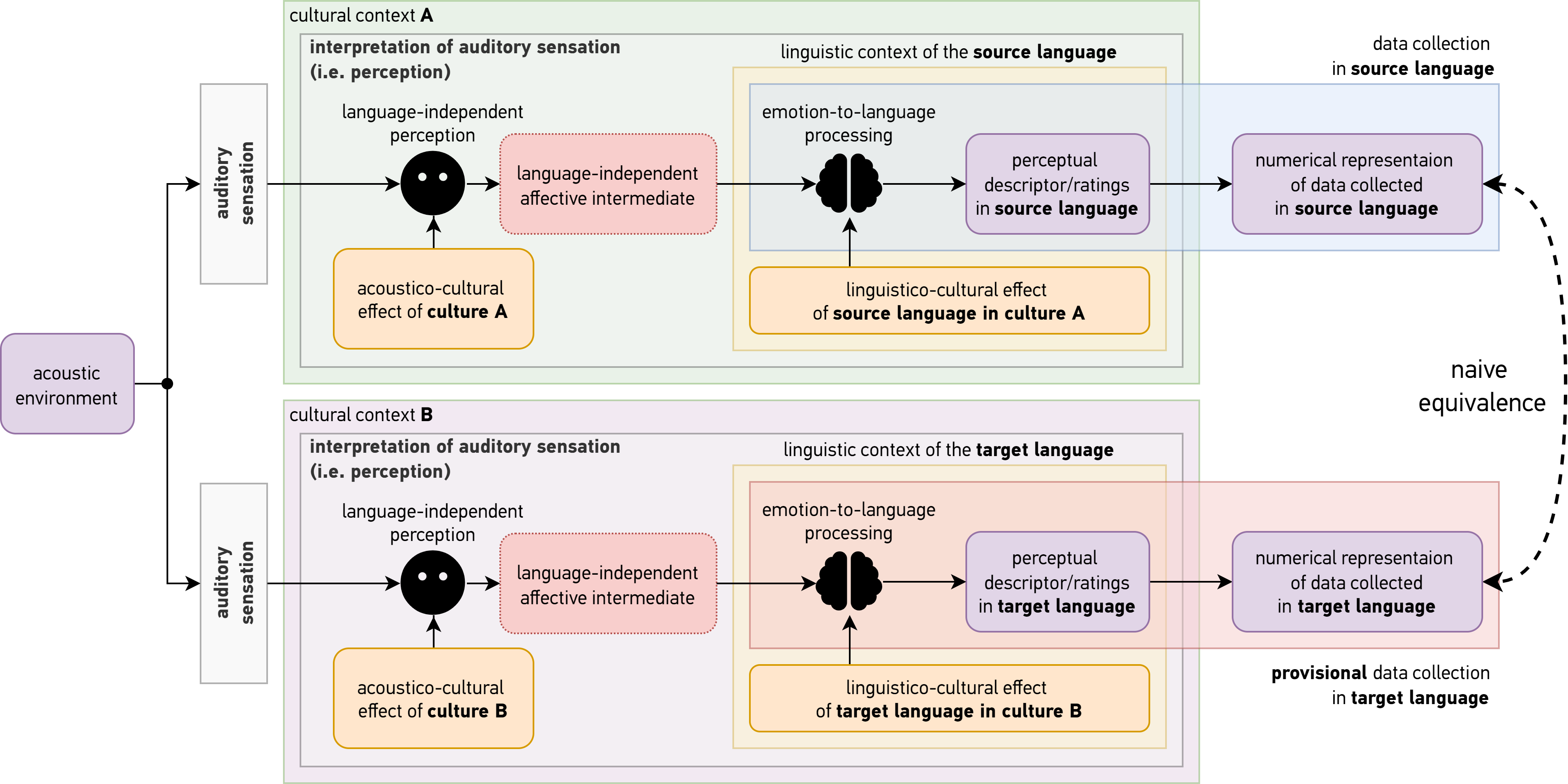}
    \caption{A multilingual model of sound perception when each language is situated in a separate cultural context. Extended from the ISO~12913-1:2014 perceptual construct of soundscape \citep[Fig.\ 1]{ISO2014ISOFramework}.}
    \label{fig:multicultural}
\end{figure*}

In a hypothetical unicultural but multilingual setting, the simplified pathway can be constructed as shown in \Cref{fig:perception}. If the data collection process (i.e., soundscape attribute translations) is developed such that the desired equivalence shown in \Cref{fig:perception} is achieved, the linguistico-cultural effect would be removed between the two languages but the acoustico-cultural effect will remain unaffected by the translation. In the real world, however, separate language-culture pairs as shown in \Cref{fig:multicultural} are often encountered. Establishing a naive equivalence between the two language-culture pairs as shown in \Cref{fig:multicultural} would result in both the acoustico- and linguistico-cultural effects being removed, hampering the ability to perform cross-cultural comparison\added{s} of soundscape perception.

On the other hand, the unicultural setting shown in \Cref{fig:perception} can often be mimicked by leveraging the availability of bilingual speakers who are fluent in both the source and target languages. In the context of most international standards, the source languages are usually one of the global \textit{lingua francas}, which are in a unique position as languages that are not nearly as intertwined to specific cultures compared to most other languages \citep[p.\ 9]{Pennycook2017TheLanguage}. Moreover, in the context of the SATP initiative where English is the source language, English tends to overwhelmingly be the second language (L2) of the translators or participants, with the target local language often being the first language (L1). As a result, the acoustico-cultural effect modulating the perception of bilingual participants is also often much closer to that of a monolingual speaker of the target language \citep{Wu2022AnType, Pavlenko2012AffectiveCognition} --- even if a multilingual person may think slightly differently in different languages \citep{Francis2005BilingualRepresentation, Emmorey2019Language:Language} --- as it is still the \textit{same} person who is perceiving. As such, differences in perceptual ratings of a stimulus on two translations of the same instrument could be attributed to a large extent to the linguistic discrepancies between the translations. In fact, the use of bilingual participants for translating and validating psychological instruments has also been suggested in guidelines such as those by \citet{Gudmundsson2009GuidelinesInstruments} and the \citet{ITC2017}.

Admittedly, a potential issue with using bilingual participants for scale validation could be due to the well-established differences in emotionality between perception in L1 and L2 \citep{Lindquist2015TheConstructionism}, which is also heavily affected by the relative proficiency between L1 and L2 \citep{Pavlenko2002BilingualismEmotions, Pavlenko2012AffectiveCognition, Panicacci2014EmotionsL2, Winskel2013TheBilinguals}. As a result, the level of English proficiency may also affect the validity of studies leveraging bilingual perception, especially if the study design utilizes a direct comparison of measurement scales. 

\subsection{Quantitative methods for soundscape translation}

\added{As noted earlier, there are no commonly-used protocols for translation of soundscape attributes. As such, most translations of both the SSQP and ISO/TS 12193 descriptors have been expert-led in a somewhat freeform manner.}
\begin{smoved}\comment{moved from \S1}
Interestingly, the Portuguese provisional translation process \citep{Antunes2021ValidatedAssessment} took a more structured and slightly more quantitative approach with Stage 1 (provisional translation) divided into two substages. Although the first substage concerning the initial selection process of candidate translations remains inevitably dependent on an expert panel, in the second substage, an online questionnaire administered cross-nationally in Portugal and Brazil was employed to finalize their set of provisional translations before proceeding to the listening test stage. For each soundscape attribute, the questionnaire item starts with a sentence describing an acoustic environment with a perceptual quality of the attribute, followed by asking which of the candidate translations is considered most ``suitable''. The participants are also given a free response box for each item to suggest another translation. The results of the questionnaire provided the Portuguese working group with quantitative insights on the suitability of each candidate translation, as well as cross-cultural differences in the use of the Portuguese language between Portugal and Brazil.
\end{smoved}

\end{sadded}

\section{Design of the Quantitative Evaluation}
\label{sec:vq}

In most translation processes, several candidate translations are often proposed via either an expert panel discussion or a parallel translation method. Traditionally, the consensus stage to obtain the final set of translations is also expert-based. However, such a consensus stage is highly qualitative and subjective in nature and provides no quantitative means of validating the experts' opinions before proceeding to subsequent validation stages. As such, the proposed quantitative evaluation framework is intended to serve as a more robust consensus-cum-bias evaluation stage of the translation process \citep{Gudmundsson2009GuidelinesInstruments, Wild2005PrinciplesAdaptation}, before proceeding to listening experiments. The design of the quantitative evaluation is based loosely on the guidelines by \citet{Gudmundsson2009GuidelinesInstruments} and the Test Development and Confirmation Guidelines in the ITC Guidelines for Translating and Adapting Tests \citep{ITC2017}. Due to the nature of the soundscape attribute translation, where individual attributes are translated to allow for standardized usage across various research methodologies, method bias cannot be investigated. As such, the main goal of the quantitative evaluation process is to assess the psychometric equivalence between the translated attributes and the standardized English attributes from the ISO 12913 series. 

Since the soundscape attributes together form a circumplex model, there is also a need to preserve the inter-attribute psychometric properties in the translations. However, testing all interactions between all local translations would be intractable. Instead, we rely on bilingual speakers of the local language and English to perform the validation. By anchoring all other attributes in English and only testing each candidate translation one at a time, it is still possible to select the most suitable candidate translation that fits in the circumplex model without using impractically many test items. This is done by ensuring that for each attribute, the interaction between the other seven English attributes and the best candidate translation is as close to that of its English counterpart as possible. By extension, it follows that the interaction between the eight final translations will approximate that of the original circumplex model in English. 

The quantitative evaluation process is questionnaire-based, with a fixed set of questionnaire items for each attribute-translation pair. To minimize any further translation, the questionnaire is conducted in English, with only the translation candidate being in the local language. \added{The use of
the intra-sentential mixed-language prompt, instead of direct pairing between perceptual attribute scales of the two languages, also encourages a certain degree of code-switching in the bilingual participants, forcing the perceptual meanings between both the source and target languages to be jointly processed and hopefully minimizing the effects of the differences in emotionality between L1 and L2. However, due to the lack of studies for this specific scenario, the extent to which this is successful will require further studies.}

Responses to each questionnaire item are rated on an 11-point Likert scale between 0 to 10. For ease of score computation, all responses are normalized to the range $[0, 1]$. We define the normalized rating of a participant $p$ by $r_{\text{[qn]}}\in[0,1]$ where $\text{[qn]}$ represent\added{s} the question, the contribution of a rating to a score by $s_{\text{[cr]}}\in[0,1]$ where $\text{[cr]}$ represent\added{s} the criterion, and the overall score by $S_\text{[cr]}\in[0,1]$. Note that all scores are purposely designed to be in the $[0,1]$ range for ease of comparison.

For brevity, we indicate a `field' in a questionnaire item in square brackets \textsc{[...]}. \textsc{[loc]} indicates the candidate translation \textit{in the local language} that is being evaluated. \textsc{[eng]} indicates a soundscape attribute \textit{in English}. Without additional qualifier, \textsc{[eng]} refers to the source attribute that \textsc{[loc]} is being translated from. \textsc{[eng]} with additional qualifier, e.g., \textsc{[adjacent~eng]}, refers to the related attribute \textit{in English} of the source attribute.

\subsection{Appropriateness (\textsc{appr})}

The first evaluation criterion for the translation is the \textit{appropriateness} of the translation with respect to the English attribute. Simply put, the translation candidate has to first be a commonly accepted translation of the English attribute with sufficiently similar meanings. \added[comment=R2.3]{By design, the appropriateness criterion can be considered a rough overall measure of linguistic equivalence and acceptability in a `lay' sense, without yet delving into the investigation of the psychometric construct. The appropriateness criterion and variations thereof are perhaps one of the most regularly used measures of translation quality for psychometric instruments both within and outside psychoacoustics. For example, in the Portuguese translation process \citep{Antunes2021ValidatedAssessment}, participants were asked to select the most \textit{suitable} translation out of the shortlisted candidates. The use of \textit{appropriateness} as a criterion perhaps dates back to at least the seminal paper by \citet{Axelsson2010APerception}. In that study, the adjectives deemed not appropriate for describing soundscapes were eliminated in the process of obtaining their final set of adjectives for describing soundscapes. Moreover, it is also important to evaluate whether the translation candidates are appropriate as a description of an acoustic environment, as highlighted in the issue of translating soundscape terms to French in \citet{Tarlao2016ComparingMontreal,Tarlao2021InvestigatingModeling}. }

The appropriateness criterion applies to all attributes on both the main and derived axes. The prompt for the appropriateness evaluation reads:
\begin{quote}
    To what extent do you agree/disagree that \textsc{[loc]} is an appropriate translation of \textsc{[eng]}?
\end{quote}
with full disagreement represented by the rating $r_\textsc{appr}=0$ and the full agreement represented by the rating $r_\textsc{appr}=1$. For example, a prompt for evaluating the appropriateness of \textit{``angenehm''} as a German translation for \textit{pleasant} would read
\begin{quote}
    To what extent do you agree/disagree that \textit{``angenehm''} is an appropriate translation of \textit{pleasant}?
\end{quote}

For \added{the} appropriateness score (\textsc{appr}), we use a simple contribution system $s_\textsc{appr} = r_\textsc{appr}$.

\subsection{Understandability (\textsc{undr})}

The next evaluation criterion is concerned with the general understandability of the translation candidate in the target population. That is, the translation candidate has to be a sufficiently commonplace term amongst the speakers of the local language, and not \deleted{an} expert or academic jargon that may not be easily understood by the general public. \added[comment=R2.3]{The need for easily understandable translations has been highlighted in the ISO/TS 12913-2:2018 standard \citep{ISO2018ISO/TSRequirements} as well as the translation guidelines by both \citet{Gudmundsson2009GuidelinesInstruments} and the \citet{ITC2017}.} This criterion applies to all attributes on both the main and derived axes. 

The prompt for the understandability evaluation reads:
\begin{quote}
    To what extent do you agree/disagree that \textsc{[loc]} is easily understood by a typical general \textsc{[local language]} speaker?
\end{quote}
with full disagreement represented by the rating $r_\textsc{undr}=0$ and the full agreement represented by the rating $r_\textsc{undr}=1$. For example, a prompt for evaluating the understandability of \textit{``dynamique''} as a French translation for \textit{vibrant} would read 
\begin{quote}
    To what extent do you agree/disagree that \textit{``dynamique''} is easily understood by a typical general \textit{French} speaker?
\end{quote}

For \added{the} understandability  score (\textsc{undr}), we also use a simple contribution $s_\textsc{undr} = r_\textsc{undr}$.

\subsection{Clarity (\textsc{clar})}

Depending on linguistic peculiarities, certain local translation candidates can be easily confused or more often associated as a translation for an adjacent attribute, instead of the target attribute. \added[comment=R2.3]{The importance of distinguishing adjacent attributes was highlighted in \citet{Jeon2018AExperiments}, where Korean translations of \textit{monotonous} and \textit{uneventful} in SSQP were found to be used almost interchangeably. A similar problem in Japanese was also raised in \citet{Nagahata2018LinguisticResearch} and experimentally confirmed in \citet{Nagahata2019ExaminationJapanese}. Translation of \textit{uneventful} to French also posed a similar problem as noted in \citet{Tarlao2016ComparingMontreal}.}
As such, it is important to quantify the degree in which the local translation candidate will be unambiguously perceived as the target translation, instead of adjacent attributes. 

The prompt for this questionnaire item reads,
\begin{quote}
    To what extent do you agree/disagree that \textsc{[loc]} is more often associated as a translation of \textsc{[\textbf{adjacent} eng]}?
\end{quote}
with full disagreement represented by the rating $r_\textsc{asso}=0$ and the full agreement represented by the rating $r_\textsc{asso}=1$.
For each attribute, clarity is evaluated twice, once against the clockwise adjacent attribute (${\curvearrowleft}$), and once against the counterclockwise adjacent attribute (${\curvearrowright}$).

For example, the two prompts for evaluating the clarity of \textit{``membosankan''} as a Bahasa Melayu translation of \textit{monotonous} would read
\begin{quote}
    To what extent do you agree/disagree that \textit{``membosankan''} is more often associated as a translation of \textit{uneventful}?
\end{quote}
and
\begin{quote}
    To what extent do you agree/disagree that \textit{``membosankan''} is more often associated as a translation of \textit{annoying}?
\end{quote}
with the rating of the former being $r_{\textsc{asso}}^{\curvearrowleft}$ as \textit{uneventful} is the counterclockwise adjacent attribute of \textit{monotonous}, and the rating of the latter being $r_{\textsc{asso}}^{\curvearrowright}$ as \textit{annoying} is the clockwise adjacent attribute of \textit{monotonous}.

The ratings from these questionnaire items are then used to compute the clarity score (\textsc{clar}), by penalizing the total extent in which the candidate translation may be confused as a translation of an adjacent attribute, such that
\begin{equation}
    s_\textsc{clar} = 1 - 0.5\left(r_{\textsc{asso}}^{\curvearrowleft} + r_{\textsc{asso}}^{\curvearrowright}\right).
\end{equation}

\subsection{Antipodal Antonymity (\textsc{anto})}

For attributes on the main axes, the translation candidates have to reflect the antonymous relationship between the pair of attributes on each end. For example, the translation of \textit{pleasant} should have an antonymous relationship to both the translation of \textit{annoying}. \added[comment=R2.3]{This criterion arises out of the need to ensure that the antipodal attributes on the main axes have affective properties that are approximately equal in magnitude and opposite in `direction', as defined in ISO/TS 12913-3:2019. In \citet{Jeon2018AExperiments}, it was found that the unvalidated translations of \textit{pleasant} and \textit{unpleasant} in SSQP for Korean do not form a proper antipodal pair across the perceptual `origin'. In the same study, it was also found that the loading of the French translation of \textit{eventful} was larger in magnitude than that of \textit{uneventful}, despite having opposite directions.} 

\deleted{However, since the VQ itself is used to evaluate the suitability of the translation candidates, we use the English term as a proxy for the evaluation of antonymity.}\added{Since the translation for the antipodal attribute itself is also not yet fixed, we use the English term for the antipodal attribute as a proxy.} The prompt for antonymity evaluation reads:
\begin{quote}
    To what extent do you agree/disagree that \textsc{[loc]} is a direct antonym of \textsc{[\textbf{antipodal} eng]}?
\end{quote}
with full disagreement represented by the rating $r_\textsc{anto}=0$ and the full agreement represented by the rating $r_\textsc{anto}=1$. 

To illustrate, the prompt for evaluating antipodal antonymity of \textit{``sinh \dj\d{\^{o}}ng''} as a Vietnamese translation of \textit{eventful} would read
\begin{quote}
    To what extent do you agree/disagree that \textit{``sinh \dj\d{\^{o}}ng''} is a direct antonym of \textit{uneventful}?
\end{quote}

\added{The antipodal}\deleted{Antipodal} antonymity score (\textsc{anto}) also uses a simple contribution system $s_\textsc{anto} = r_\textsc{anto}$.

\subsection{Orthogonal Unbiasedness (\textsc{orth})}

For attributes on the main axes, it is also important that the linguistic orthogonality between the pleasant-annoying and the eventful-uneventful axes are preserved after the translation. In other words, an attribute on the main axis should be as neutral as possible with respect to the two orthogonal attributes on the other main axis. \added[comment=R2.3]{The importance of orthogonal unbiasedness was again demonstrated in the trinational study \citep{Jeon2018AExperiments} on the SSQP, where skewing of the main axes was found in all three languages studied.}

As with antonymity, we use the English terms as a proxy for the evaluation. The prompt for bias evaluation reads:
\begin{quote}
    To what extent is \textsc{[loc]} (as a description of an acoustic environment) biased with respect to the \textsc{[\textbf{ccw orthogonal} eng]}--\textsc{[\textbf{cw orthogonal} eng]} axis?
\end{quote}
where full bias towards the clockwise orthogonal attribute is represented by the rating $r_\textsc{bias}=0$ and full bias towards the counterclockwise orthogonal attribute is represented by the rating $r_\textsc{bias}=1$.  For instance, the prompt for evaluating the orthogonal bias of \textit{``keyifsiz''} as a Turkish translation of \textit{annoying} would read
\begin{quote}
    To what extent is \textit{``keyifsiz''} (as a description of an acoustic environment) biased with respect to the \textit{uneventful}--\textit{eventful} axis?
\end{quote}
with full bias towards \textit{uneventful} represented by the rating $r_\textsc{bias}=0$ and full bias towards \textit{eventful} represented by the rating $r_\textsc{bias}=1$.

The orthogonality score (\textsc{orth}) is based on the extent which the rating deviates from the neutral point ($r_\textsc{bias}=0.5$), and is thus given by
\begin{equation}
    s_\textsc{orth} = 1 - 2\left\lvert r_\textsc{bias} - 0.5\right\rvert.
\end{equation}

\subsection{Connotativeness (\textsc{conn}), Nonconnotativeness (\textsc{ncon}), and Implicative Balance (\textsc{ibal})}

Lastly, the implicative meanings of the translated candidates are evaluated. For this particular category, the desired behavior of the candidate translation differs between attributes on the main axes and those on the derived axes. For attributes on the main axes, their translation\added{s} should also preserve their `basis' nature, in the sense that they should not imply adjacent attributes on the derived axes. For attributes on the derived axes, their translations should imply both adjacent attributes on the main axes. To illustrate, describing a soundscape as \textit{pleasant} does not imply that the soundscape is necessarily \textit{calm} nor \textit{vibrant}. On the other hand, describing a soundscape as \textit{vibrant}, should imply that the soundscape is both \textit{pleasant} and \textit{eventful}. Additionally, since the attributes on the derived axes also serve as the approximate anchor of the angular midpoint between their respective adjacent attributes on the main axes, translations of the attributes on the derived axes should serve the same function in the target language. \added[comment=R2.3]{In both \citet{Tarlao2021InvestigatingModeling} and \citet{Jeon2018AExperiments}, nearly all SSQP equivalents of the derived-axis attributes in ISO 12913 were found to suffer from significant imbalances in loadings across the languages studied.} Note the distinction between the concept of \textit{implication} in this item and the concept of \textit{confusion} in the \textsc{clar} criterion.

The questionnaire prompt for this item reads,
\begin{quote}
    To what extent do you agree/disagree that \textsc{[loc]} (as a description of an acoustic environment) implies that the environment is also \textsc{[\textbf{adjacent} eng]}?
\end{quote}
with full disagreement represented by the rating $r_\textsc{impl}=0$ and the full agreement represented by the rating $r_\textsc{impl}=1$. 

For example, the two questionnaire prompts to obtain the implicative ratings \textit{``kaotiskt''} as a Swedish translation of \textit{chaotic} would read
\begin{quote}
    To what extent do you agree/disagree that \textit{``kaotiskt''} (as a description of an acoustic environment) implies that the environment is also \textit{annoying}?
\end{quote}
and
\begin{quote}
    To what extent do you agree/disagree that \textit{``kaotiskt''} (as a description of an acoustic environment) implies that the environment is also \textit{eventful}?
\end{quote}
with the former represented by $r_{\textsc{impl}}^{ \curvearrowleft}$ as \textit{annoying} is the counterclockwise adjacent attribute of \textit{chaotic}, and the latter by $r_{\textsc{impl}}^{ \curvearrowright}$ as \textit{eventful} is the clockwise adjacent attribute of \textit{chaotic}.

For attribute\added{s} on the main axes, where implying adjacent attributes is undesirable, the non-connotativeness score (\textsc{ncon}) is computed similarly to the clarity score, such that
\begin{equation}
    s_{\textsc{ncon}} = 1 - 0.5 \left(r_{\textsc{impl}}^{ \curvearrowleft} + r_{\textsc{impl}}^{\curvearrowright}\right).
\end{equation}

For attributes on the derived axes, where implying adjacent attributes is desirable, the connotativeness score (\textsc{conn}) is given by
\begin{equation}
    s_{\textsc{conn}} = 0.5 \left(r_{\textsc{impl}}^{ \curvearrowleft} + r_{\textsc{impl}}^{\curvearrowright}\right).
\end{equation}

Lastly, the implicative balance score (\textsc{ibal}) is computed by penalizing the difference between the two $r_\textsc{impl}$ scores, such that
\begin{equation}
    s_\textsc{ibal} = 1 - \left|r_{\textsc{impl}}^{ \curvearrowleft}-r_{\textsc{impl}}^{\curvearrowright}\right|.
\end{equation}

Although the \textsc{ibal} score is mainly designed to evaluate attributes on the derived axes, it is important to note that in practice, no translation of the attributes on the main axes would be perfectly non-connotative of their respective adjacent attributes. With this in mind, we also compute the \textsc{ibal} score for attributes on the main axes to ensure that, even if they are not completely non-connotative of adjacent attributes, the extent of connotativeness remains similar between the clockwise adjacent and the counterclockwise adjacent.

\section{Phase I: Initial Translation by Experts}\label{sec:init}

\subsection{Methodology}

Due to the lack of a translation protocol specific to psychoacoustics, we relied on several guidelines used for translation of psychological instruments \citep{Gudmundsson2009GuidelinesInstruments, Borsa2012Cross-culturalConsiderations, ITC2017} and made adjustments specific to the requirements of soundscape attribute translation.

\subsubsection{Translation team}

The initial expert translation phase involved five linguistic experts who are all native Thai speakers and bilingual in English. As the linguists are not soundscape experts, summaries of the soundscape methodologies and standards, particularly on the circumplex model, were provided prior to the start of the translation process. The translation process is additionally facilitated by a soundscape researcher who is also a native Thai speaker. 

\subsubsection{Method of translation}

The experts were first asked to independently produce a set of potential translations for each of the eight English soundscape attributes \textit{without} consulting any other expert. As remarked in \citet{Gudmundsson2009GuidelinesInstruments}, a process based on parallel independent translations was adopted as the translation methodology, instead of validation via back translation, due to the need to prioritize the psychometric equivalence of the translated soundscape attributes. Moreover, the use of independent translations has the additional benefit of helping to identify as much potential ambiguity in translation as possible. 

Due to the constraints associated with the translations of the soundscape attributes, the experts \deleted{are}\added{were} also provided with guidelines specific to the English-to-Thai translation of soundscape attributes, which \deleted{was}\added{were} adapted from the generic provisional translation guidelines used in the SATP project. The guidelines provided are summarized in the next section (\Cref{sssec:guidelines}).

Following the initial translation, the experts were convened to discuss the initially\added{-}proposed translations. New candidate translations may also be proposed at this point. Instead of the usual aim of reaching a consensus for a \textit{final} translation, the goal of this discussion \deleted{is}\added{was} to shortlist a few candidate translations per soundscape attribute to proceed to the quantitative evaluation phase. By doing so, the typically subjective and unquantifiable consensus process for arriving at a final translation set is eliminated and replaced by quantitative analysis.

The shortlisted translation candidates at the end of this phase are shown in \Cref{tab:candidates}. 

\begin{table*}[t]
    \caption{Translation candidates for soundscape attributes after Stage 1 (Expert Translation). Asterisk (*) indicates a translation candidate that \deleted{has been}\added{was} proposed for multiple English attributes.}
    \label{tab:candidates}
    \renewcommand*{\arraystretch}{1.1}
    \centering
    \renewcommand{\tabcolsep}{3pt}
    \begin{tabularx}{\textwidth}{l@{\hskip 6pt}l@{\hskip 3pt}l *{3}{l@{\hskip 3pt}l}}
    \toprule
    Eng. Attr. & \multicolumn{8}{@{\hskip 3pt}l}{Thai Translation Candidates} \\
    \midrule
Pleasant	&	\thainaafang{} & \naafang& \thairuenhuu{} & \ruenhuu& \thaisanorhuu{} & \sanorhuu\\
Annoying	&	\thainaaramkaan{} & \naaramkaan& \thairakaaihuu{} & \rakaaihuu\\
Eventful	&	\thaimiiarai{} & \miiarai& \thaiwunwaai{}* & \wunwaai& \thaiuekgatuek{} & \uekgatuek\\
Uneventful	&	\thaisangop{}* & \sangop& \thairiiap{} & \riiap& \thaimaimiiarai{} & \maimiiarai\\
\midrule
Calm	&	\thaipornklaai{} & \pornklaai& \thaisangop{}* & \sangop& \thaisabaaihuu{} & \sabaaihuu\\
Chaotic	&	\thaijorjae{} & \jorjae& \thaibpanbpuuan{} & \bpanbpuuan& \thaiyungyerng{} & \yungyerng& \thaiwunwaai{}* & \wunwaai\\
Monotonous	&	\thaijuedjued{} & \juedjued& \thainaabuea{} & \naabuea& \thainuuainuuai{} & \nuuainuuai& \thaiuuaiuuai{} & \uuaiuuai\\
Vibrant	&	\thaikuekkak{} & \kuekkak& \thaimiichiiwitchiiwaa{} & \miichiiwitchiiwaa& \thaisotsai{} & \sotsai& \thaiwuewaa{} & \wuewaa\\
    \bottomrule
    \end{tabularx}
\end{table*}

\subsubsection{Translation guidelines} \label{sssec:guidelines}

First, the translations should strive to use common words that are easily understood by laypeople. This also means academic terms and jargon which may not be easily understood by most Thai-speaking general population should be avoided. 

Next, each English attribute should be translated in relation to the perception of sounds. It is more desirable to retain the \textit{meaning} in the acoustic-perceptual sense rather than pursuing a literal translation. If a single Thai word does not sufficiently capture the original meaning of the English attribute, a set of two to three Thai words can be proposed instead. The use of \thaiwali{} \wali{}, \lit{short phrases}, are also allowed since the distinction between adjectives and an adjectival phrases can be ambiguous in Thai \citep{Post2008AdjectivesClasses}.

For the attributes on the main axes, the attributes should be translated as neutrally as possible with respect to the orthogonal axis. For example, \textit{pleasant} and \textit{annoying} should be translated as neutrally with respect to the \textit{eventful}-\textit{uneventful} axis as possible. Additionally, it is desirable that translations of antipodal attributes are antonyms of each other. However, as far as possible, \added{the experts were advised to} avoid the use of the negative particle \thaimai{} \mai{}, \lit{not}, to \replaced{minimize}{avoid} ambiguity. Similar to the English word `\textit{not}', \thaimai{} \mai{} may be interpreted either as a true negating operator --- e.g. \textit{not cold} being interpreted as \textit{warm} --- or as a neutralizing operator --- e.g. \textit{not cold} being interpreted as \textit{neither cold nor warm} \citep{Atlas1977NegationPresupposition, Takahashi1997NegationStudy}.

\subsection{Discussion}

\subsubsection{The use of auditory indicator morpheme in candidate translations}

Several candidate translations were proposed by the experts to contain the morpheme \thaihuu{} \huu, \lit{ear}, to indicate that the adjective is acoustical in nature. The morphemes preceding \thaihuu{} \huu{} in most translation candidates containing it do not inherently have acoustical connotations, but the addition of \thaihuu{} \huu{} provides a clear indication that the preceding morpheme is describing an acoustic environment. For example, \thairuen{} \ruen{}, \lit{enjoyable or comfortable}, in \thairuenhuu{} \ruenhuu{} and \thairakaai{} \rakaai{}, \lit{to irritate}, in \thairakaaihuu{} \rakaaihuu{}. In a similar manner, \thaifang{} \fang{}, \lit{to listen}, serves a similar purpose in \thainaafang{} \naafang{}.

\subsubsection{Translation of \textit{eventful} and \textit{uneventful}}

Amongst the perceptual attributes, the so-called arousal-axis attributes \textit{eventful} and \textit{uneventful} prove\added{d} particularly difficult to translate, especially with respect to the connotation of valency \added{--- an issue similarly encountered in the translation process for French \citep{Tarlao2016ComparingMontreal}, Korean \citep{Jeon2018AExperiments}, and Japanese \citep{Nagahata2018LinguisticResearch, Nagahata2019ExaminationJapanese}.} Translations of \textit{eventful} and \textit{uneventful} to Thai usually depend on the context and typically take the form of one of the adjacent attributes on the derived axes. For this pair of attributes, the guideline on the avoidance of \thaimai{} \mai{}, \lit{not}, was relaxed, partly due to the presence of the `\textit{un-}' morpheme in \textit{uneventful}. 

A particularly interesting pair of translation candidates are \thaimiiarai{} \miiarai{} and \thaimaimiiarai{} \maimiiarai{}, as the pair are morphologically very close to the English terms. The Thai morpheme \thaimii{} \mii, \lit{to have}, can be considered to correspond to the English morpheme `\textit{-ful}', while \thaiarai{} \arai, \lit{something}, can be considered to correspond to the free morpheme \mbox{`\textit{-event-}}'. Although both terms were remarked by the experts as not commonly used in formal Thai, it was nonetheless agreed by them as perhaps the most inherently neutral with respect to perceptual valency out of all the eventful-uneventful candidates. Other candidate translations of \textit{eventful} and \textit{uneventful} were all considered by at least one expert to be inherently leaning towards one of the terms on the derived axes during the discussion -- an opinion which will be later supported by the evaluation questionnaire.

\section{Phase II: Quantitative Evaluation on Thai Translation Candidates}
\label{sec:quant}

The quantitative evaluation phase was conducted between November 2021 and February 2022, during which a total of 31 participants who are bilingual in Thai and English were recruited across Thailand, Singapore, and the United Kingdom. The participants were also asked to report their length of stay outside Thailand, of which 4 participants (\SI{12.9}{\percent}) reported less than one year, 13 participants (\SI{41.9}{\percent}) reported between 1 to 5 years, 6 participants (\SI{19.4}{\percent}) reported between 6 to 10 years, and 8 participants (\SI{25.8}{\percent}) reported more than 10 years.

The participants were asked to rate their proficiency in Thai and English on the Interagency Language Roundtable (ILR) scale. For the Thai language, 30 participants (\SI{96.8}{\percent}) reported Native Proficiency and 1 participant (\SI{3.2}{\percent}) reported Professional Working Proficiency. For English, 5 participants (\SI{16.1}{\percent}) reported Native Proficiency (ILR~5), 12 participants (\SI{38.7}{\percent}) reported Full Professional Proficiency (ILR~4), 9 participants (\SI{29.0}{\percent}) reported Professional Working Proficiency (ILR~3), 4 participants (\SI{12.9}{\percent}) reported Limited Working Proficiency (ILR~2), and 1 participant (\SI{3.2}{\percent}) reported Elementary Proficiency (ILR~1).

The survey was administered online via Google Forms. Google Apps Script was used to programmatically generate the questionnaire. In total, the questionnaire contains 178 items, excluding the demographic information section.

\subsection{Results}

The evaluation scores introduced in \Cref{sec:vq} were calculated using the results. For each evaluation criterion of each English attribute, without assuming normality, a Kruskal–-Wallis test \citep[KWT; ][]{Kruskal1952UseAnalysis} was performed on the score contributions with respect to the candidate translations. If a statistical significance was found at \SI{5}{\percent} significance level using the KWT, then a posthoc Conover--Iman test\footnote{The Conover--Iman test is similar to the more well-known Dunn's test \citep{Dunn1964MultipleSums} but uses the Student's \textit{t}-distribution instead of the normal distribution. We use the more statistically powerful Conover--Iman test in this paper.} \citep[CIT;][]{Conover1979OnProcedures} with Bonferroni correction \citep{Dunn1961MultipleMeans} was performed to identify pairwise differences. No posthoc test is performed if the KWT is not statistically significant.

\Cref{tab:main_scores} and \Cref{tab:derived_scores} show the mean evaluation scores\footnote{We have also experimented with weighting the contributions to the scores by the sum of English and Thai proficiency ILR rating, but the weighting has negligible effects on the scores.}. The results of the pairwise tests are shown in \Cref{tab:citp} in the Appendix. 

The raw results of the questionnaire are shown in \Cref{fig:rawmain} and \Cref{fig:rawderived} for the attributes on the main and derived axes, respectively. The following sections will discuss the results of the validation questionnaire and the final translation candidates, which are also summarized in \Cref{tab:finals}.

% The radar plots of the mean evaluation scores are also shown in \Cref{fig:main_radar} and \Cref{fig:derived_radar}.
\newcommand{\tablegend}{Double asterisks (**) and single asterisk (*) with boldface indicate that the distribution underlying the score is significantly different from those of \textbf{all} other candidates, at \SI{1}{\percent} and \SI{5}{\percent} significance levels using the pairwise posthoc Conover-Iman test with Bonferroni corrections, respectively. For candidates with at least one but not all statistically significant posthoc tests, the oplus ($^{\oplus}$) sign indicates the number of pairwise tests where the null hypotheses are rejected at \SI{1}{\percent} significance level. The plus ($^{+}$) sign indicates the number of pairwise tests where the null hypotheses are rejected at \SI{5}{\percent} significance level but not at \SI{1}{\percent}.}

\begin{table*}[t]
    \caption{Evaluation scores for attributes on the main axes. \tablegend{}}
    \label{tab:main_scores}
    \renewcommand*{\arraystretch}{1.1}
    \centering
    \begin{tabularx}{\textwidth}{lll*{7}{>{\raggedleft\arraybackslash}X}}
    \toprule
    Eng. Attr. &
    \multicolumn{2}{l}{Thai Translation Candidate} &
    {\uppercase{appr}} &
    {\uppercase{undr}} &
    {\uppercase{clar}} &
    {\uppercase{anto}} &
    {\uppercase{orth}} &
    {\uppercase{ncon}} &
    {\uppercase{ibal}} \\
    \midrule Pleasant & \thainaafang & \naafang & 0.868 & **\textbf{0.974} & 0.560 & 0.677 & 0.632 & 0.503 & 0.768 \\
 & \thaisanorhuu & \sanorhuu & 0.832 & 0.732 & 0.516 & 0.687 & 0.755 & 0.479 & 0.803 \\
 & \thairuenhuu & \ruenhuu & 0.858 & $^{+}${0.842} & 0.532 & 0.681 & 0.697 & 0.482 & 0.758 \\
\midrule Annoying & \thainaaramkaan & \naaramkaan & **\textbf{0.974} & **\textbf{0.984} & 0.574 & 0.823 & 0.619 & 0.542 & 0.645 \\
 & \thairakaaihuu & \rakaaihuu & 0.810 & 0.819 & 0.635 & 0.771 & 0.652 & 0.626 & 0.645 \\
\midrule Eventful & \thaimiiarai & \miiarai & 0.297 & $^{+}${0.729} & **\textbf{0.737} & 0.426 & $^{\oplus}${0.729} & **\textbf{0.718} & $^{\oplus}${0.810} \\
 & \thaiuekgatuek & \uekgatuek & $^{+}${0.474} & 0.635 & 0.555 & 0.487 & $^{+}${0.535} & 0.482 & $^{\oplus}${0.687} \\
 & \thaiwunwaai & \wunwaai & 0.416 & **\textbf{0.945} & 0.429 & 0.474 & 0.323 & 0.394 & 0.445 \\
\midrule Uneventful & \thaimaimiiarai & \maimiiarai & $^{\oplus}${0.745} & $^{+}${0.923} & $^{\oplus}${0.492} & 0.694 & *\textbf{0.832} & $^{+}${0.415} & $^{\oplus}${0.823} \\
 & \thaisangop & \sangop & 0.690 & $^{\oplus}${0.961} & 0.345 & $^{+}${0.729} & 0.239 & 0.319 & 0.458 \\
 & \thairiiap & \riiap & 0.548 & 0.781 & 0.373 & 0.587 & $^{\oplus}${0.665} & 0.331 & $^{\oplus}${0.835} \\
    \bottomrule 
    \end{tabularx}
\end{table*}

\subsubsection{Pleasant}

With the exception of \textsc{undr}, all translation candidates for \textit{pleasant} performed similarly across all other criteria, showing no significant difference using the KWT at \SI{5}{\percent}. Since \thainaafang{} \naafang{} performed significantly better than all other translation candidates in terms of understandability, with CIT ${p<0.001}$ against all other candidates, \thainaafang{} \naafang{} was selected as the final translation for \textit{pleasant}.

\subsubsection{Annoying}

Both candidates for \textit{annoying} show no statistically significant difference across the \textsc{clar}, \textsc{anto}, \textsc{orth}, \textsc{ncon}, and \textsc{ibal} criteria. For the \textsc{appr} and \textsc{undr} criteria, \thainaaramkaan{} \naaramkaan{} scored higher than \thairakaaihuu{} \rakaaihuu{} with CIT ${p<0.001}$ on both. As such, \thainaaramkaan{} \naaramkaan{} was selected as the final translation for \textit{annoying}.
 
\subsubsection{Eventful}

With the exception of \textsc{anto}, KWT showed statistically significant differences in all other criteria for \textit{eventful}. 

In terms of \textsc{appr}, \thaiuekgatuek{} \uekgatuek{} performed better than \thaimiiarai{} \miiarai{} with \eventfulAPPRuekgatuekGTmiiarai{}. Other pairs show no significant difference at \SI{5}{\percent}. In terms of \textsc{undr}, \thaiwunwaai{} \wunwaai{} performed better than \thaimiiarai{} \miiarai{} and \thaiuekgatuek{} \uekgatuek{}, with ${p<0.001}$ on both, while \thaimiiarai{} \miiarai{} performed better than \thaiuekgatuek{} \uekgatuek{} with \eventfulUNDRmiiaraiGTuekgatuek{}. For the remaining criteria, \thaimiiarai{} \miiarai{} performed significantly better than \thaiwunwaai{} \wunwaai{} with ${p<0.001}$ in all. Against \thaiuekgatuek{} \uekgatuek{}, \thaimiiarai{} \miiarai{} performed better in \textsc{clar} and \textsc{ncon}, but no significant difference was found in \textsc{orth} and \textsc{ibal}.

As earlier remarked by the experts during the initial translation phase, neither is \thaimiiarai{} \miiarai{} a word commonly used for the translation of eventful, nor is it a word commonly found in formal usage of the Thai language. As such, the lower \textsc{appr} and \textsc{undr} scores are somewhat expected. In fact, \thaimiiarai{} \miiarai{} being an uncommon word may have contributed to the high \textsc{clar} score, as it would not be easily confused as a translation of another soundscape descriptor. \thaiwunwaai{} \wunwaai{}, however, despite having a very high \textsc{undr} score, is very easily confused as a translation of \textit{chaotic} as seen in \Cref{fig:rawmain} --- and it, in fact, will be chosen as the final translation for \textit{chaotic} in \Cref{sssec:chaotic}. Considering the evaluation criteria as a whole, \thaimiiarai{} \miiarai{} was selected as the final translation for \textit{eventful}.

\subsubsection{Uneventful}

In terms of \textsc{appr}, \thaimaimiiarai{} \maimiiarai{} performed significantly better than \thairiiap{} \riiap{} with \uneventfulAPPRmaimiiaraiGTriiap{}, but no significant difference was found against \thaisangop{} \sangop{}. In terms of \textsc{undr}, \thaisangop{} \sangop{} and \thaimaimiiarai{} \maimiiarai{} both performed better than \thairiiap{} \riiap{}, with ${p<0.001}$ and \uneventfulUNDRmaimiiaraiGTriiap{}, respectively. No significant difference was found between \thaisangop{} \sangop{} and \thaimaimiiarai{} \maimiiarai{}. Interestingly, despite being just as uncommonly used as the \textit{eventful} candidate \thaimiiarai{} \miiarai{}, \thaimaimiiarai{} \maimiiarai{} was rated with relatively good \textsc{appr} ($\mu=0.745$) and \textsc{undr} ($\mu=0.923$) scores, whereas the former was rated much lower with \textsc{appr} ($\mu=0.297$) and \textsc{undr} ($\mu=0.729$).

In terms of \textsc{anto}, \thaisangop{} \sangop{} performed better than \thairiiap{} \riiap{} with \uneventfulANTOsangopGTriiap{}, but no significant differences were found with other pairs. With \textsc{orth}, \thaimaimiiarai{} \maimiiarai{} performed better than both other candidates with ${p<0.001}$ against \thaisangop{} \sangop{} and \uneventfulORTHmaimiiaraiGTriiap{} against \thairiiap{} \riiap{}. At the same time, \thairiiap{} \riiap{} performed better than \thaisangop{} \sangop{} with ${p<0.001}$. As shown in \Cref{fig:rawmain}, \thaisangop{} \sangop{} is generally rated as strongly biased towards pleasant, and a similar bias is also seen with \thairiiap{} \riiap{} to a lesser degree.

In terms of \textsc{clar}, \thaimaimiiarai{} \maimiiarai{} performs better than \thaisangop{} \sangop{} with ${p<0.001}$, but significant difference was not found against \thairiiap{} \riiap{} with \uneventfulCLARmaimiiaraiGTriiap{} at \SI{5}{\percent} significance level. \thairiiap{} \riiap{} and \thaisangop{} \sangop{} performed very similarly in this criterion, with ${p\approx1}$, although the nature of the association is somewhat different. \thaisangop{} \sangop{} has a very strong association as a translation of \textit{calm} rather than \textit{uneventful} but much less so with \textit{monotonous}. On the other hand, \thairiiap{} \riiap{} has moderately high associations as a translation of both \textit{calm} and \textit{monotonous}. A similar result was seen with \textsc{ncon}, where \thaimaimiiarai{} \maimiiarai{} performed better than \thaisangop{} \sangop{} with \uneventfulNCONmaimiiaraiGTsangop{}, but no significant difference was found against \thairiiap{} \riiap{}. Expectedly, analysis of the \textsc{ibal} scores also shows that both \thaimaimiiarai{} \maimiiarai{} and \thairiiap{} \riiap{} performed significantly better than \thaisangop{} \sangop{} with ${p<0.001}$. 

Considering all criteria as a whole, \thaimaimiiarai{} \maimiiarai{} was chosen as the final translation, particularly due to the poor evaluation of \thaisangop{} \sangop{} in \textsc{clar}, \textsc{orth}, and \textsc{ncon}.

\begin{table*}[t]
\caption{Evaluation scores for attributes on the derived axes. \tablegend{}}
    \label{tab:derived_scores}
    \renewcommand*{\arraystretch}{1.1}
    \centering
    \begin{tabularx}{\textwidth}{lll*{5}{>{\raggedleft\arraybackslash}X}}
    \toprule
    Eng. Attr. &
    \multicolumn{2}{l}{Thai Translation Candidate} &
    {\uppercase{appr}} &
    {\uppercase{undr}} &
    {\uppercase{clar}} &
    {\uppercase{conn}} & 
    {\uppercase{ibal}}\\
    \midrule Calm & \thaisangop & \sangop & **\textbf{0.958} & $^{\oplus}${0.977} & 0.394 & 0.697 & $^{\oplus}${0.774} \\
 & \thaipornklaai & \pornklaai & $^{\oplus}${0.745} & $^{+}${0.942} & 0.468 & 0.624 & 0.674 \\
 & \thaisabaaihuu & \sabaaihuu & 0.568 & 0.871 & 0.445 & 0.581 & 0.516 \\
\midrule Chaotic & \thaiwunwaai & \wunwaai & $^{\oplus}${0.919} & *\textbf{0.955} & 0.371 & 0.679 & 0.739 \\
 & \thaiyungyerng & \yungyerng & $^{\oplus}${0.868} & $^{\oplus}${0.877} & $^{+}${0.494} & 0.618 & 0.797 \\
 & \thaibpanbpuuan & \bpanbpuuan & $^{\oplus}${0.842} & 0.852 & 0.476 & 0.595 & 0.823 \\
 & \thaijorjae & \jorjae & 0.697 & 0.748 & 0.474 & 0.619 & 0.787 \\
\midrule Vibrant & \thaimiichiiwitchiiwaa & \miichiiwitchiiwaa & $^{\oplus\oplus}${0.932} & $^{\oplus}${0.929} & 0.277 & $^{\oplus}${0.763} & $^{+}${0.848} \\
 & \thaisotsai & \sotsai & $^{\oplus}${0.803} & $^{\oplus}${0.935} & 0.377 & $^{\oplus}${0.669} & 0.752 \\
 & \thaiwuewaa & \wuewaa & 0.539 & 0.710 & **\textbf{0.573} & 0.510 & 0.748 \\
 & \thaikuekkak & \kuekkak & $^{\oplus}${0.877} & $^{\oplus}${0.913} & 0.292 & $^{\oplus}${0.755} & 0.697 \\
\midrule Monotonous & \thainaabuea & \naabuea & 0.716 & **\textbf{0.968} & 0.448 & 0.552 & 0.639 \\
 & \thaiuuaiuuai & \uuaiuuai & 0.652 & 0.787 & 0.527 & 0.513 & 0.619 \\
 & \thainuuainuuai & \nuuainuuai & 0.674 & 0.719 & 0.532 & 0.502 & 0.681 \\
 & \thaijuedjued & \juedjued & 0.732 & $^{+}${0.839} & 0.545 & 0.474 & 0.568 \\
    \bottomrule 
    \end{tabularx}
\end{table*}

\subsubsection{Calm} \label{sssec:calm}

Whereas \thaisangop{} \sangop{} did not perform well as a translation of \textit{uneventful}, it was rated better as a translation of \textit{calm}. In terms of \textsc{appr}, \thaisangop{} \sangop{} performed better than both other candidates with ${p<0.001}$. \thaisangop{} \sangop{} also performed better than \thaisabaaihuu{} \sabaaihuu{} in \textsc{undr} and \textsc{ibal} with ${p<0.001}$.
For \textsc{clar} and \textsc{conn}, no significant difference in distribution were found using the omnibus test. As such, \thaisangop{} \sangop{} was selected as the final translation for \textit{calm}.

\subsubsection{Chaotic} \label{sssec:chaotic}

With \textsc{appr}, all other candidates performed better than \thaijorjae{} \jorjae{}, with ${p<0.001}$ for both \thaiwunwaai{} \wunwaai{} and \thaiyungyerng{} \yungyerng{}, and \chaoticAPPRbpanbpuuanGTjorjae{} for \thaibpanbpuuan{} \bpanbpuuan{}. Excluding \thaijorjae{} \jorjae{}, no other significant pairwise differences were found for \textsc{appr}. 

For \textsc{undr}, \thaiwunwaai{} \wunwaai{} performed better than all other candidates with ${p<0.001}$ against \thaijorjae{} \jorjae{} and \thaibpanbpuuan{} \bpanbpuuan{}, and \chaoticUNDRwunwaaiGTyungyerng{} against \thaiyungyerng{} \yungyerng{}. For \textsc{clar}, the only significant pairwise difference \added{was} with \thaiyungyerng{} \yungyerng{} against \thaiwunwaai{} \wunwaai{} with \chaoticCLARyungyerngGTwunwaai. No significant differences in distribution were found for \textsc{conn} and \textsc{ibal}. 

% {\color{red} 
Considering that \thaiwunwaai{} \wunwaai{} outperformed all other candidates in \textsc{undr} and performed similarly to \thaiyungyerng{} \yungyerng{} and \thaibpanbpuuan{} \bpanbpuuan{} in \textsc{appr}, \thaiwunwaai{} \wunwaai{} was selected as the final translation for \textit{chaotic}.
% }

\subsubsection{Vibrant}

The omnibus tests indicate statistically significant differences in distributions for all evaluation criteria. For \textsc{appr}, \thaimiichiiwitchiiwaa{} \miichiiwitchiiwaa{} performed better than \thaisotsai{} \sotsai{} with \vibrantAPPRmiichiiwitchiiwaaGTsotsai{} and \thaiwuewaa{} \wuewaa{} with ${p<0.001}$. \thaisotsai{} \sotsai{} and \thaikuekkak{} \kuekkak{} also performed better than \thaiwuewaa{} \wuewaa{} with ${p<0.001}$. \thaimiichiiwitchiiwaa{} \miichiiwitchiiwaa{} and \thaikuekkak{} \kuekkak{} did not have statistically significant difference\added{s} (\vibrantAPPRmiichiiwitchiiwaaGTkuekkak{}), and neither did \thaisotsai{} \sotsai{} and \thaikuekkak{} \kuekkak{} (\vibrantAPPRkuekkakGTsotsai{}).

In terms of \textsc{undr}, \thaimiichiiwitchiiwaa{} \miichiiwitchiiwaa{}, \thaisotsai{} \sotsai{}, and \thaikuekkak{} \kuekkak{} all performed similarly with ${p\approx1}$. Against \thaiwuewaa{} \wuewaa{}, the former three all performed better with ${p<0.001}$. Interestingly, despite the relatively lower appropriateness and understandability ratings, \thaiwuewaa{} \wuewaa{} performed the best in terms of \textsc{clar} with ${p<0.001}$ against the other three, while other pairs are not statistically significant. The better performance of \thaiwuewaa{} \wuewaa{} is likely due to the other three having a stronger association towards \textit{pleasantness} than they do \textit{annoying}, as seen in \Cref{fig:rawderived}. 

However, \thaiwuewaa{} \wuewaa{} performed poorly in terms of \textsc{conn}, as it indicates \textit{eventfulness} much more strongly than it does \textit{pleasantness}. The posthoc tests give ${p<0.001}$ with respect to \thaikuekkak{} \kuekkak{} and \thaimiichiiwitchiiwaa{} \miichiiwitchiiwaa{}, and \vibrantCONNsotsaiGTwuewaa{} with respect to \thaisotsai{} \sotsai{}. In terms of \textsc{ibal}, \thaisotsai{} \sotsai{}, \thaikuekkak{} \kuekkak{}, and \thaiwuewaa{} \wuewaa{} performed very similarly with ${p\approx1}$. \thaimiichiiwitchiiwaa{} \miichiiwitchiiwaa{} performed better than \thaikuekkak{} \kuekkak{} with \vibrantIBALmiichiiwitchiiwaaGTkuekkak{}. All other pairwise tests are not statistically significant at \SI{5}{\percent}.

Overall, \thaimiichiiwitchiiwaa{} \miichiiwitchiiwaa{} performed the strongest in \textsc{undr} and \textsc{ibal}, with no statistically significant differences against \thaikuekkak{} \kuekkak{} and \thaisotsai{} \sotsai{} in \textsc{clar} and \textsc{conn}. As such, \thaimiichiiwitchiiwaa{} \miichiiwitchiiwaa{} was selected as the final translation for \textit{vibrant}.
 
\subsubsection{Monotonous}

\comment{R1.1}

Except for \textsc{undr}, no other criterion shows statistically significant differences in performance. Since \thainaabuea{} \naabuea{} outperformed the rest in terms of \textsc{undr} with ${p<0.001}$, \thainaabuea{} \naabuea{} was selected as the final translation for \textit{monotonous}.

\begin{table}[t]
\caption{Final Thai translation of the circumplex soundscape attributes.}
    \label{tab:finals}
\renewcommand*{\arraystretch}{1.1}
    \centering
    \begin{tabularx}{\columnwidth}{XlX}
    \toprule
    English Attribute & \multicolumn{2}{l}{Final Thai Translation} \\
    \midrule
    Pleasant	&	\thainaafang{}              & \naafang\\
    Annoying	&	\thainaaramkaan{}           & \naaramkaan\\
    Eventful	&	\thaimiiarai{}              & \miiarai\\
    Uneventful	&	\thaimaimiiarai{}           & \maimiiarai\\
    \midrule
    Calm	    &	\thaisangop{}               & \sangop\\
    Chaotic	    &	\thaiwunwaai{}              & \wunwaai\\
    Monotonous	&	\thainaabuea{}              & \naabuea\\
    Vibrant	    &	\thaimiichiiwitchiiwaa{}    & \miichiiwitchiiwaa\\
    \bottomrule
    \end{tabularx}
\end{table}

\section{Discussion and Conclusion}\label{sec:conclusion}

In this work, we proposed a structured and quantitative framework for the translation of soundscape attributes from the standardized English terms in the ISO 12913 series of standards to a local language. By considering the inter-attribute relationships of soundscape attributes on the circumplex model, a set of questionnaire items and corresponding evaluation criteria was developed to evaluate the linguistic-cultural suitability and psychometric equivalence of the candidate translations. The proposed framework was then applied to the soundscape attribute translation process for the Thai language as an initial study. 

The translation process involves two phases, the first being parallel translations by linguistic experts and a group discussion to obtain a shortlist of candidate translations. The second phase applied the proposed quantitative framework to assess the translation candidates and select the final set of translations based on the evaluation scores. In total, 31 participants who are bilingual in Thai and English were recruited to participate in the pilot validation questionnaire. The proposed evaluation metrics were computed based on the raw questionnaire responses and statistical tests were performed to identify the most suitable translation for each of the soundscape attributes.

The use of a quantitative framework has greatly facilitated the process of identifying and verifying the strengths and weaknesses of each candidate translation, and the data can continue to provide insights into the linguistic and psychometric properties of the translations for the experimental validation stage and future studies based on this set of translations. 

\begin{sadded}
% It must be kept in mind, however, that the framework remains situated in the linguistic and cultural context of the target language. In several languages, it has been noted that there is no single `perfect' translation for some of the soundscape attributes \citep{Nagahata2018LinguisticResearch, Nagahata2019ExaminationJapanese, Tarlao2016ComparingMontreal, Jeon2018AExperiments, Aletta2020SoundscapeLanguages}. 

The data collected under the quantitative framework can also be used as conditioning or calibrating data for future studies. For example, the ratings or evaluation scores can be used as additional measurement terms in numerical techniques, such as structural equation modelling \citep[SEM; see][]{Ullman2012StructuralModeling} and confirmatory factor analysis \citep[CFA; see][]{Brown2012ConfirmatoryAnalysis}, to better compensate for known construct deviations of the translated versions. To illustrate, the Montreal study in \citet{Tarlao2021InvestigatingModeling} may benefit from having the evaluation scores for the French translations available to assist the modelling, or SEM findings from the Korean study in \citet{Hong2015InfluenceApproach} can be transformed for use in other settings with the evaluation data for Korean translations.

As with any survey and statistical method, a sufficiently large sample representative of the target population is required. However, due to the reliance on bilingual speakers, the cultural equivalence between the monolingual speakers of the target language and the bilingual speakers is crucial in ensuring the validity of the quantitative framework. As such, having multiple intermediary languages, instead of relying solely on English, may also be useful in ensuring multilingual equivalence across all translations. For example, regional \textit{lingua francas}, such as Hindi, Spanish, French, or Arabic, can be used as alternate intermediary languages to allow access to a larger pool of bilingual speakers. It is also possible to adapt the framework for comparison of any two languages which may not necessarily include\deleted{s} English. Additionally, establishing a bilingual relation between two sets of translations allows comparison to other languages whose relationship to one of the languages in previously studied pairs has been established. 

\end{sadded}

The proposed quantitative framework is also currently being used for the translation of soundscape attributes to Bahasa Melayu, which is a \added{national} language \replaced{of}{used in} both Malaysia and Singapore. The quantitative nature of the assessment data would allow for statistical comparison of cross-national differences and similarities in the linguistic and psychometric properties of the candidate translations. Although the proposed framework has been constructed specifically for the ISO 12913 circumplex model, it is also applicable to other octant-based circumplex models. Moreover, it is also possible to adapt or extend the current framework for the translation of other psychoacoustic descriptors or models where the preservation of the inter-descriptor construct is crucial. 

All in all, the authors hope that the proposed framework will continue to assist other researchers in the field of soundscapes and psychoacoustics working on a translation, and act as a step forward in transforming the traditionally subjective and heavily expert-reliant process into one that is more robust and verifiable.

\section*{Declaration of competing interest}
The authors declare that they have no known competing financial interests or personal relationships that could have appeared to influence the work reported in this paper.

\section*{Acknowledgments}
This work was supported by the Google Cloud Research Credits program (GCP205559654).

The authors would like to thank Dr.\@ Francesco Aletta, Dr.\@ Tin Oberman, Andrew Mitchell, and Prof.\@ Jian Kang, of the UCL Institute for Environmental Design and Engineering, The Bartlett Faculty of the Built Environment, University College London (UCL), London, United Kingdom, for coordinating the SATP project and providing assistance for the Thai Language Working Group. 

We would like to also thank 
Pulaporn Sreewichian (School of Computing, Engineering and Built Environment, Glasgow Caledonian University, Glasgow, Scotland, United Kingdom),
Kan Jitpakdi (Thai Student Society, Singapore; National University of Singapore, Singapore), 
Phumrapee Pisutsin (Thai Student Society, Singapore; Nanyang Technological University, Singapore), 
Nerinat Yongphiphatwong (Samaggi Samagom, United Kingdom; University of Cambridge, Cambridge, United Kingdom), and 
Siddha Kumwongwan (Samaggi Samagom, United Kingdom; University College London, London, United Kingdom), 
for their assistance with the distribution of the quantitative validation questionnaire. 

\printcredits

\bibliographystyle{cas-model2-names}
% \bibliographystyle{unsrt}
% \balance
\bibliography{refs,iso}

\begin{thebibliography}{54}
\expandafter\ifx\csname natexlab\endcsname\relax\def\natexlab#1{#1}\fi
\providecommand{\url}[1]{\texttt{#1}}
\providecommand{\href}[2]{#2}
\providecommand{\path}[1]{#1}
\providecommand{\DOIprefix}{doi:}
\providecommand{\ArXivprefix}{arXiv:}
\providecommand{\URLprefix}{URL: }
\providecommand{\Pubmedprefix}{pmid:}
\providecommand{\doi}[1]{\href{http://dx.doi.org/#1}{\path{#1}}}
\providecommand{\Pubmed}[1]{\href{pmid:#1}{\path{#1}}}
\providecommand{\bibinfo}[2]{#2}
\ifx\xfnm\relax \def\xfnm[#1]{\unskip,\space#1}\fi
%Type = Inproceedings
\bibitem[{Aletta et~al.(2020)Aletta, Oberman, Axelsson, Xie, Zhang, Lau, Tang,
  Jambro{\v{s}}ic, de~Coensel, van~den Bosch, Aumond, Guastavino, Lavandier,
  Fiebig, Schulte-Fortkamp, Sarwono, Sudarsono, Astolfi, Nagahata, Jeon, Jo,
  Chieng, Gan, Hong, Lam, Ong, Kogan, Silva, Manzano, Y{\"{o}}r{\"{u}}koglu,
  Nguyen and Kang}]{Aletta2020SoundscapeLanguages}
\bibinfo{author}{Aletta, F.}, \bibinfo{author}{Oberman, T.},
  \bibinfo{author}{Axelsson, O.}, \bibinfo{author}{Xie, H.},
  \bibinfo{author}{Zhang, Y.}, \bibinfo{author}{Lau, S.K.},
  \bibinfo{author}{Tang, S.K.}, \bibinfo{author}{Jambro{\v{s}}ic, K.},
  \bibinfo{author}{de~Coensel, B.}, \bibinfo{author}{van~den Bosch, K.},
  \bibinfo{author}{Aumond, P.}, \bibinfo{author}{Guastavino, C.},
  \bibinfo{author}{Lavandier, C.}, \bibinfo{author}{Fiebig, A.},
  \bibinfo{author}{Schulte-Fortkamp, B.}, \bibinfo{author}{Sarwono, J.},
  \bibinfo{author}{Sudarsono, A.}, \bibinfo{author}{Astolfi, A.},
  \bibinfo{author}{Nagahata, K.}, \bibinfo{author}{Jeon, J.Y.},
  \bibinfo{author}{Jo, H.I.}, \bibinfo{author}{Chieng, J.},
  \bibinfo{author}{Gan, W.S.}, \bibinfo{author}{Hong, J.Y.},
  \bibinfo{author}{Lam, B.}, \bibinfo{author}{Ong, Z.T.},
  \bibinfo{author}{Kogan, P.}, \bibinfo{author}{Silva, E.S.},
  \bibinfo{author}{Manzano, J.V.}, \bibinfo{author}{Y{\"{o}}r{\"{u}}koglu,
  P.N.D.}, \bibinfo{author}{Nguyen, T.L.}, \bibinfo{author}{Kang, J.},
  \bibinfo{year}{2020}.
\newblock \bibinfo{title}{{Soundscape assessment: Towards a validated
  translation of perceptual attributes in different languages}}, in:
  \bibinfo{booktitle}{Proceedings of the 49th International Congress and Expo
  on Noise Control Engineering}.
%Type = Inproceedings
\bibitem[{Antunes et~al.(2021)Antunes, Michalski, de~Ulh{\^{o}}a~Carvalho and
  Alves}]{Antunes2021ValidatedAssessment}
\bibinfo{author}{Antunes, S.}, \bibinfo{author}{Michalski, R.L.X.N.},
  \bibinfo{author}{de~Ulh{\^{o}}a~Carvalho, M.L.}, \bibinfo{author}{Alves, S.},
  \bibinfo{year}{2021}.
\newblock \bibinfo{title}{{Validated translation into Portuguese of perceptual
  attributes for soundscape assessment}}, in: \bibinfo{booktitle}{Proceedings
  of the 12th European Congress and Exposition on Noise Control Engineering},
  pp. \bibinfo{pages}{710--718}.
%Type = Article
\bibitem[{Atlas(1977)}]{Atlas1977NegationPresupposition}
\bibinfo{author}{Atlas, J.D.}, \bibinfo{year}{1977}.
\newblock \bibinfo{title}{{Negation, ambiguity, and presupposition}}.
\newblock \bibinfo{journal}{Linguistics and Philosophy} \bibinfo{volume}{1},
  \bibinfo{pages}{321--336}.
\newblock \DOIprefix\doi{10.1007/BF00353452}.
%Type = Inproceedings
\bibitem[{Axelsson et~al.(2009)Axelsson, Nilsson and
  Berglund}]{Axelsson2009AQuality}
\bibinfo{author}{Axelsson, O.}, \bibinfo{author}{Nilsson, M.E.},
  \bibinfo{author}{Berglund, B.}, \bibinfo{year}{2009}.
\newblock \bibinfo{title}{{A Swedish instrument for measuring soundscape
  quality}}, in: \bibinfo{booktitle}{Proceedings of the 8th European Conference
  on Noise Control}.
%Type = Article
\bibitem[{Axelsson et~al.(2010)Axelsson, Nilsson and
  Berglund}]{Axelsson2010APerception}
\bibinfo{author}{Axelsson, O.}, \bibinfo{author}{Nilsson, M.E.},
  \bibinfo{author}{Berglund, B.}, \bibinfo{year}{2010}.
\newblock \bibinfo{title}{{A principal components model of soundscape
  perception}}.
\newblock \bibinfo{journal}{Journal of the Acoustical Society of America}
  \bibinfo{volume}{128}, \bibinfo{pages}{2836--2846}.
\newblock \DOIprefix\doi{10.1121/1.3493436}.
%Type = Article
\bibitem[{Axelsson et~al.(2012)Axelsson, Nilsson and
  Berglund}]{Axelsson2012TheProtocol}
\bibinfo{author}{Axelsson, O.}, \bibinfo{author}{Nilsson, M.E.},
  \bibinfo{author}{Berglund, B.}, \bibinfo{year}{2012}.
\newblock \bibinfo{title}{{The Swedish soundscape-quality protocol}}.
\newblock \bibinfo{journal}{Journal of the Acoustical Society of America}
  \bibinfo{volume}{131}, \bibinfo{pages}{3476}.
\newblock \URLprefix \url{https://doi.org/10.1121/1.4709112},
  \DOIprefix\doi{10.1121/1.4709112}.
%Type = Article
\bibitem[{Borsa et~al.(2012)Borsa, Dam{\'{a}}sio and
  Bandeira}]{Borsa2012Cross-culturalConsiderations}
\bibinfo{author}{Borsa, J.C.}, \bibinfo{author}{Dam{\'{a}}sio, B.F.},
  \bibinfo{author}{Bandeira, D.R.}, \bibinfo{year}{2012}.
\newblock \bibinfo{title}{{Cross-cultural adaptation and validation of
  psychological instruments: Some considerations}}.
\newblock \bibinfo{journal}{Paid{\'{e}}ia (Ribeir{\~{a}}o Preto)}
  \bibinfo{volume}{22}, \bibinfo{pages}{423--432}.
%Type = Article
\bibitem[{Brown and Moore(2012)}]{Brown2012ConfirmatoryAnalysis}
\bibinfo{author}{Brown, T.A.}, \bibinfo{author}{Moore, M.T.},
  \bibinfo{year}{2012}.
\newblock \bibinfo{title}{{Confirmatory factor analysis}}.
\newblock \bibinfo{journal}{Handbook of structural equation modeling}
  \bibinfo{volume}{361}, \bibinfo{pages}{379}.
%Type = Article
\bibitem[{Cain et~al.(2013)Cain, Jennings and Poxon}]{Cain2013TheSoundscape}
\bibinfo{author}{Cain, R.}, \bibinfo{author}{Jennings, P.},
  \bibinfo{author}{Poxon, J.}, \bibinfo{year}{2013}.
\newblock \bibinfo{title}{{The development and application of the emotional
  dimensions of a soundscape}}.
\newblock \bibinfo{journal}{Applied Acoustics} \bibinfo{volume}{74},
  \bibinfo{pages}{232--239}.
\newblock \URLprefix \url{http://dx.doi.org/10.1016/j.apacoust.2011.11.006},
  \DOIprefix\doi{10.1016/j.apacoust.2011.11.006}.
%Type = Techreport
\bibitem[{Conover and Iman(1979)}]{Conover1979OnProcedures}
\bibinfo{author}{Conover, W.J.}, \bibinfo{author}{Iman, R.L.},
  \bibinfo{year}{1979}.
\newblock \bibinfo{title}{{On multiple-comparisons procedures}}.
\newblock \bibinfo{type}{Technical Report}. Los Alamos Scientific Laboratory.
%Type = Article
\bibitem[{Deng et~al.(2020)Deng, Kang, Zhao and
  Jambro{\v{s}}i{\'{c}}}]{Deng2020Cross-NationalCroatia}
\bibinfo{author}{Deng, L.}, \bibinfo{author}{Kang, J.}, \bibinfo{author}{Zhao,
  W.}, \bibinfo{author}{Jambro{\v{s}}i{\'{c}}, K.}, \bibinfo{year}{2020}.
\newblock \bibinfo{title}{{Cross-National Comparison of Soundscape in Urban
  Public Open Spaces between China and Croatia}}.
\newblock \bibinfo{journal}{Applied Sciences} \bibinfo{volume}{10},
  \bibinfo{pages}{960}.
%Type = Article
\bibitem[{Dunn(1961)}]{Dunn1961MultipleMeans}
\bibinfo{author}{Dunn, O.J.}, \bibinfo{year}{1961}.
\newblock \bibinfo{title}{{Multiple Comparisons Among Means}}.
\newblock \bibinfo{journal}{Journal of the American Statistical Association}
  \bibinfo{volume}{56}, \bibinfo{pages}{52}.
\newblock \DOIprefix\doi{10.2307/2282330}.
%Type = Article
\bibitem[{Dunn(1964)}]{Dunn1964MultipleSums}
\bibinfo{author}{Dunn, O.J.}, \bibinfo{year}{1964}.
\newblock \bibinfo{title}{{Multiple Comparisons Using Rank Sums}}.
\newblock \bibinfo{journal}{Technometrics} \bibinfo{volume}{6},
  \bibinfo{pages}{241--252}.
\newblock \URLprefix \url{http://www.jstor.org/stable/1266041},
  \DOIprefix\doi{10.2307/1266041}.
%Type = Article
\bibitem[{Emmorey(2019)}]{Emmorey2019Language:Language}
\bibinfo{author}{Emmorey, K.}, \bibinfo{year}{2019}.
\newblock \bibinfo{title}{{Language: Do Bilinguals Think Differently in Each
  Language?}}
\newblock \bibinfo{journal}{Current Biology} \bibinfo{volume}{29},
  \bibinfo{pages}{R1133--R1135}.
\newblock \URLprefix \url{https://doi.org/10.1016/j.cub.2019.09.009},
  \DOIprefix\doi{10.1016/j.cub.2019.09.009}.
%Type = Article
\bibitem[{Francis(2005)}]{Francis2005BilingualRepresentation}
\bibinfo{author}{Francis, W.S.}, \bibinfo{year}{2005}.
\newblock \bibinfo{title}{{Bilingual semantic and conceptual representation}}.
\newblock \bibinfo{journal}{Handbook of bilingualism: Psycholinguistic
  approaches} , \bibinfo{pages}{251--267}\URLprefix
  \url{http://books.google.com/books?hl=en&amp;lr=&amp;id=2fzBDptA5NMC&amp;oi=fnd&amp;pg=PA251&amp;dq=Bilingual+Semantic+and+Conceptual+Representation&amp;ots=Pn2pGeYLip&amp;sig=OSNzgks8cnFph-bzEK-Gd18HSEs}.
%Type = Incollection
\bibitem[{Goddard(2015)}]{Goddard2015WordsMeaning}
\bibinfo{author}{Goddard, C.}, \bibinfo{year}{2015}.
\newblock \bibinfo{title}{{Words as Carriers of Cultural Meaning}}, in:
  \bibinfo{booktitle}{The Oxford Handbook of the Word}.
\newblock \DOIprefix\doi{10.1163/9789004357723}.
%Type = Article
\bibitem[{Gudmundsson(2009)}]{Gudmundsson2009GuidelinesInstruments}
\bibinfo{author}{Gudmundsson, E.}, \bibinfo{year}{2009}.
\newblock \bibinfo{title}{{Guidelines for translating and adapting
  psychological instruments}}.
\newblock \bibinfo{journal}{Nordic Psychology} \bibinfo{volume}{61},
  \bibinfo{pages}{29--45}.
\newblock \DOIprefix\doi{10.1027/1901-2276.61.2.29}.
%Type = Article
\bibitem[{Hansen and Weber(2009)}]{Hansen2009SemanticComparison}
\bibinfo{author}{Hansen, H.}, \bibinfo{author}{Weber, R.},
  \bibinfo{year}{2009}.
\newblock \bibinfo{title}{{Semantic evaluations of noise with tonal components
  in Japan, France, and Germany: A cross-cultural comparison}}.
\newblock \bibinfo{journal}{Journal of the Acoustical Society of America}
  \bibinfo{volume}{125}, \bibinfo{pages}{850--862}.
\newblock \URLprefix \url{https://doi.org/10.1121/1.3050275},
  \DOIprefix\doi{10.1121/1.3050275}.
%Type = Article
\bibitem[{Harzing et~al.(2002)Harzing, Maznevski, Fischlmayr, Yaconi,
  Wittenberg, Myloni, Kong~Low, Castro, Zander, Karlsson, Romani and
  Feely}]{Harzing2002TheCountries}
\bibinfo{author}{Harzing, A.W.}, \bibinfo{author}{Maznevski, M.},
  \bibinfo{author}{Fischlmayr, I.}, \bibinfo{author}{Yaconi, L.L.},
  \bibinfo{author}{Wittenberg, K.}, \bibinfo{author}{Myloni, B.},
  \bibinfo{author}{Kong~Low, J.C.}, \bibinfo{author}{Castro, F.B.},
  \bibinfo{author}{Zander, L.}, \bibinfo{author}{Karlsson, C.},
  \bibinfo{author}{Romani, L.}, \bibinfo{author}{Feely, A.},
  \bibinfo{year}{2002}.
\newblock \bibinfo{title}{{The interaction between language and culture: A test
  of the cultural accommodation hypothesis in seven countries}}.
\newblock \bibinfo{journal}{Language and Intercultural Communication}
  \bibinfo{volume}{2}, \bibinfo{pages}{120--139}.
\newblock \DOIprefix\doi{10.1080/14708470208668081}.
%Type = Article
\bibitem[{Hong and Jeon(2015)}]{Hong2015InfluenceApproach}
\bibinfo{author}{Hong, J.Y.}, \bibinfo{author}{Jeon, J.Y.},
  \bibinfo{year}{2015}.
\newblock \bibinfo{title}{{Influence of urban contexts on soundscape
  perceptions: A structural equation modeling approach}}.
\newblock \bibinfo{journal}{Landscape and Urban Planning}
  \bibinfo{volume}{141}, \bibinfo{pages}{78--87}.
\newblock \URLprefix \url{http://dx.doi.org/10.1016/j.landurbplan.2015.05.004},
  \DOIprefix\doi{10.1016/j.landurbplan.2015.05.004}.
%Type = Inproceedings
\bibitem[{Huang and Knapp(2018)}]{Huang2018AnMusic}
\bibinfo{author}{Huang, W.}, \bibinfo{author}{Knapp, R.B.},
  \bibinfo{year}{2018}.
\newblock \bibinfo{title}{{An exploratory study of population differences based
  on massive database of physiological responses to music}}, in:
  \bibinfo{booktitle}{Proceedings of the 7th International Conference on
  Affective Computing and Intelligent Interaction}, pp.
  \bibinfo{pages}{524--530}.
\newblock \DOIprefix\doi{10.1109/ACII.2017.8273649}.
%Type = Misc
\bibitem[{{\acroauthor{International Organization for
  Standardization}{ISO}}(2014)}]{ISO2014ISOFramework}
\bibinfo{author}{{\acroauthor{International Organization for
  Standardization}{ISO}}}, \bibinfo{year}{2014}.
\newblock \bibinfo{title}{{ISO 12913-1 Acoustics. Soundscape Part 1: Definition
  and conceptual framework}}.
%Type = Misc
\bibitem[{{\acroauthor{International Organization for
  Standardization}{ISO}}(2018)}]{ISO2018ISO/TSRequirements}
\bibinfo{author}{{\acroauthor{International Organization for
  Standardization}{ISO}}}, \bibinfo{year}{2018}.
\newblock \bibinfo{title}{{ISO/TS 12913-2 Acoustics. Soundscape Part 2: Data
  collection and reporting requirements}}.
%Type = Misc
\bibitem[{{\acroauthor{International Organization for
  Standardization}{ISO}}(2019)}]{ISO2019ISO/TSAnalysis}
\bibinfo{author}{{\acroauthor{International Organization for
  Standardization}{ISO}}}, \bibinfo{year}{2019}.
\newblock \bibinfo{title}{{ISO/TS 12913-3 Acoustics. Soundscape Part 3: Data
  analysis}}.
%Type = Misc
\bibitem[{{\acroauthor{International Test Commission}{ITC}}(2017)}]{ITC2017}
\bibinfo{author}{{\acroauthor{International Test Commission}{ITC}}},
  \bibinfo{year}{2017}.
\newblock \bibinfo{title}{{The ITC Guidelines for Translating and Adapting
  Tests (Second edition)}}.
%Type = Article
\bibitem[{Jeon et~al.(2018)Jeon, Hong, Lavandier, Lafon, Axelsson and
  Hurtig}]{Jeon2018AExperiments}
\bibinfo{author}{Jeon, J.Y.}, \bibinfo{author}{Hong, J.Y.},
  \bibinfo{author}{Lavandier, C.}, \bibinfo{author}{Lafon, J.},
  \bibinfo{author}{Axelsson, O.}, \bibinfo{author}{Hurtig, M.},
  \bibinfo{year}{2018}.
\newblock \bibinfo{title}{{A cross-national comparison in assessment of urban
  park soundscapes in France, Korea, and Sweden through laboratory
  experiments}}.
\newblock \bibinfo{journal}{Applied Acoustics} \bibinfo{volume}{133},
  \bibinfo{pages}{107--117}.
\newblock \DOIprefix\doi{10.1016/j.apacoust.2017.12.016}.
%Type = Article
\bibitem[{Juslin et~al.(2016)Juslin, Barradas, Ovsiannikow, Limmo and
  Thompson}]{Juslin2016PrevalenceCultures}
\bibinfo{author}{Juslin, P.N.}, \bibinfo{author}{Barradas, G.T.},
  \bibinfo{author}{Ovsiannikow, M.}, \bibinfo{author}{Limmo, J.},
  \bibinfo{author}{Thompson, W.F.}, \bibinfo{year}{2016}.
\newblock \bibinfo{title}{{Prevalence of emotions, mechanisms, and motives in
  music listening: A comparison of individualist and collectivist cultures}}.
\newblock \bibinfo{journal}{Psychomusicology: Music, Mind, and Brain}
  \bibinfo{volume}{26}, \bibinfo{pages}{293--326}.
\newblock \DOIprefix\doi{10.1037/pmu0000161}.
%Type = Article
\bibitem[{Kotz and Paulmann(2011)}]{Kotz2011EmotionBrain}
\bibinfo{author}{Kotz, S.A.}, \bibinfo{author}{Paulmann, S.},
  \bibinfo{year}{2011}.
\newblock \bibinfo{title}{{Emotion, Language, and the Brain}}.
\newblock \bibinfo{journal}{Language and Linguistics Compass}
  \bibinfo{volume}{5}, \bibinfo{pages}{108--125}.
%Type = Article
\bibitem[{Kruskal and Wallis(1952)}]{Kruskal1952UseAnalysis}
\bibinfo{author}{Kruskal, W.H.}, \bibinfo{author}{Wallis, W.A.},
  \bibinfo{year}{1952}.
\newblock \bibinfo{title}{{Use of Ranks in One-Criterion Variance Analysis}}.
\newblock \bibinfo{journal}{Journal of the American Statistical Association}
  \bibinfo{volume}{47}, \bibinfo{pages}{583--621}.
\newblock \URLprefix \url{http://www.jstor.org/stable/2280779},
  \DOIprefix\doi{10.2307/2280779}.
%Type = Article
\bibitem[{Kuwano et~al.(1999)Kuwano, Namba, Florentine, Da~Rui, Fastl and
  Schick}]{Kuwano1999ANoise}
\bibinfo{author}{Kuwano, S.}, \bibinfo{author}{Namba, S.},
  \bibinfo{author}{Florentine, M.}, \bibinfo{author}{Da~Rui, Z.},
  \bibinfo{author}{Fastl, H.}, \bibinfo{author}{Schick, A.},
  \bibinfo{year}{1999}.
\newblock \bibinfo{title}{{A cross-cultural study of the factors of sound
  quality of environmental noise}}.
\newblock \bibinfo{journal}{Journal of the Acoustical Society of America}
  \bibinfo{volume}{105}, \bibinfo{pages}{1081}.
\newblock \URLprefix \url{https://doi.org/10.1121/1.424845},
  \DOIprefix\doi{10.1121/1.424845}.
%Type = Article
\bibitem[{Lindquist et~al.(2015)Lindquist, MacCormack and
  Shablack}]{Lindquist2015TheConstructionism}
\bibinfo{author}{Lindquist, K.A.}, \bibinfo{author}{MacCormack, J.K.},
  \bibinfo{author}{Shablack, H.}, \bibinfo{year}{2015}.
\newblock \bibinfo{title}{{The role of language in emotion: Predictions from
  psychological constructionism}}.
\newblock \bibinfo{journal}{Frontiers in Psychology} \bibinfo{volume}{6},
  \bibinfo{pages}{1--17}.
\newblock \DOIprefix\doi{10.3389/fpsyg.2015.00444}.
%Type = Article
\bibitem[{Mohamed and Dokmeci~Yorukoglu(2020)}]{Mohamed2020IndoorTurkey}
\bibinfo{author}{Mohamed, M.A.E.}, \bibinfo{author}{Dokmeci~Yorukoglu, P.N.},
  \bibinfo{year}{2020}.
\newblock \bibinfo{title}{{Indoor soundscape perception in residential spaces:
  A cross-cultural analysis in Ankara, Turkey}}.
\newblock \bibinfo{journal}{Building Acoustics} \bibinfo{volume}{27},
  \bibinfo{pages}{35--46}.
\newblock \DOIprefix\doi{10.1177/1351010X19885030}.
%Type = Book
\bibitem[{Nacaskul(2013)}]{Nacaskul2013TheThai}
\bibinfo{author}{Nacaskul, K.}, \bibinfo{year}{2013}.
\newblock \bibinfo{title}{{The Phonological System in Thai}}.
\newblock \bibinfo{edition}{7th} ed., \bibinfo{publisher}{Chulalongkorn
  University Press}, \bibinfo{address}{Bangkok}.
%Type = Inproceedings
\bibitem[{Nagahata(2018)}]{Nagahata2018LinguisticResearch}
\bibinfo{author}{Nagahata, K.}, \bibinfo{year}{2018}.
\newblock \bibinfo{title}{{Linguistic issues we must resolve before the
  standardization of soundscape research}}, in: \bibinfo{booktitle}{Proceedings
  of the 11th European Congress and Exposition on Noise Control Engineering},
  pp. \bibinfo{pages}{2459--2464}.
%Type = Article
\bibitem[{Nagahata(2019)}]{Nagahata2019ExaminationJapanese}
\bibinfo{author}{Nagahata, K.}, \bibinfo{year}{2019}.
\newblock \bibinfo{title}{{Examination of soundscape-quality protocols in
  Japanese}}.
\newblock \bibinfo{journal}{Proceedings of the 48th International Congress and
  Exhibition on Noise Control Engineering} .
%Type = Article
\bibitem[{Panicacci(2014)}]{Panicacci2014EmotionsL2}
\bibinfo{author}{Panicacci, A.}, \bibinfo{year}{2014}.
\newblock \bibinfo{title}{{Emotions from a bilingual point of view: personality
  and emotional intelligence in relation to perception and expression of
  emotions in the L1 and L2}}.
\newblock \bibinfo{journal}{International Journal of Bilingual Education and
  Bilingualism} \bibinfo{volume}{17}, \bibinfo{pages}{727--730}.
\newblock \DOIprefix\doi{10.1080/13670050.2013.857505}.
%Type = Article
\bibitem[{Pavlenko(2002)}]{Pavlenko2002BilingualismEmotions}
\bibinfo{author}{Pavlenko, A.}, \bibinfo{year}{2002}.
\newblock \bibinfo{title}{{Bilingualism and emotions}}.
\newblock \bibinfo{journal}{Multilingua} \bibinfo{volume}{21},
  \bibinfo{pages}{45--78}.
\newblock \DOIprefix\doi{10.1515/mult.2002.004}.
%Type = Article
\bibitem[{Pavlenko(2012)}]{Pavlenko2012AffectiveCognition}
\bibinfo{author}{Pavlenko, A.}, \bibinfo{year}{2012}.
\newblock \bibinfo{title}{{Affective processing in bilingual speakers:
  Disembodied cognition?}}
\newblock \bibinfo{journal}{International Journal of Psychology}
  \bibinfo{volume}{47}, \bibinfo{pages}{405--428}.
\newblock \DOIprefix\doi{10.1080/00207594.2012.743665}.
%Type = Book
\bibitem[{Pennycook and Candlin(2017)}]{Pennycook2017TheLanguage}
\bibinfo{author}{Pennycook, A.}, \bibinfo{author}{Candlin, C.N.},
  \bibinfo{year}{2017}.
\newblock \bibinfo{title}{{The cultural politics of English as an international
  language}}.
\newblock \bibinfo{publisher}{Routledge}.
%Type = Inproceedings
\bibitem[{Phan et~al.(2008)Phan, Nishimura, Phan, Yano, Sato and
  Hashimoto}]{Phan2008AnnoyanceJapanese}
\bibinfo{author}{Phan, H.A.T.}, \bibinfo{author}{Nishimura, T.},
  \bibinfo{author}{Phan, H.Y.T.}, \bibinfo{author}{Yano, T.},
  \bibinfo{author}{Sato, T.}, \bibinfo{author}{Hashimoto, Y.},
  \bibinfo{year}{2008}.
\newblock \bibinfo{title}{{Annoyance from road traffic noise with horn sounds:
  A cross-cultural experiment between Vietnamese and Japanese}}, in:
  \bibinfo{booktitle}{Proceedings of the 9th Congress of the International
  Commission on the Biological Effects of Noise}, pp.
  \bibinfo{pages}{688--698}.
%Type = Article
\bibitem[{Post(2008)}]{Post2008AdjectivesClasses}
\bibinfo{author}{Post, M.}, \bibinfo{year}{2008}.
\newblock \bibinfo{title}{{Adjectives in Thai: Implications for a functionalist
  typology of word classes}}.
\newblock \bibinfo{journal}{Linguistic Typology} \bibinfo{volume}{12},
  \bibinfo{pages}{339--381}.
\newblock \DOIprefix\doi{10.1515/LITY.2008.041}.
%Type = Article
\bibitem[{Schatz et~al.(2012)Schatz, Egger-Lampl and
  Masuch}]{Schatz2012TheRatings}
\bibinfo{author}{Schatz, R.}, \bibinfo{author}{Egger-Lampl, S.},
  \bibinfo{author}{Masuch, K.}, \bibinfo{year}{2012}.
\newblock \bibinfo{title}{{The Impact of Test Duration on User Fatigue and
  Reliability of Subjective Quality Ratings}}.
\newblock \bibinfo{journal}{Journal of the Audio Engineering Society}
  \bibinfo{volume}{60}, \bibinfo{pages}{63--73}.
%Type = Article
\bibitem[{Schwarz et~al.(2016)Schwarz, Lemaitre, Aramaki and
  Kronland-Martinet}]{Schwarz2016EffectsTests}
\bibinfo{author}{Schwarz, D.}, \bibinfo{author}{Lemaitre, G.},
  \bibinfo{author}{Aramaki, M.}, \bibinfo{author}{Kronland-Martinet, R.},
  \bibinfo{year}{2016}.
\newblock \bibinfo{title}{{Effects of test duration in subjective listening
  tests}}.
\newblock \bibinfo{journal}{Proceedings of the 42nd International Computer
  Music Conference} , \bibinfo{pages}{514--518}.
%Type = Article
\bibitem[{Sperber et~al.(1994)Sperber, Devellis and
  Boehlecke}]{Sperber1994Cross-culturalValidation}
\bibinfo{author}{Sperber, A.D.}, \bibinfo{author}{Devellis, R.F.},
  \bibinfo{author}{Boehlecke, B.}, \bibinfo{year}{1994}.
\newblock \bibinfo{title}{{Cross-cultural translation: Methodology and
  Validation}}.
\newblock \bibinfo{journal}{Journal of Cross-Cultural Psychology}
  \bibinfo{volume}{25}, \bibinfo{pages}{501--524}.
\newblock \DOIprefix\doi{10.1177/0022022194254006}.
%Type = Article
\bibitem[{Sudarsono et~al.(2022)Sudarsono, Setiasari, Sarwono and
  Nitidara}]{Sudarsono2022TheStudy}
\bibinfo{author}{Sudarsono, A.S.}, \bibinfo{author}{Setiasari, W.},
  \bibinfo{author}{Sarwono, S.J.}, \bibinfo{author}{Nitidara, N.P.A.},
  \bibinfo{year}{2022}.
\newblock \bibinfo{title}{{The development of standard perceptual attributes in
  Indonesian for soundscape evaluation: Result from initial study}}.
\newblock \bibinfo{journal}{Journal of Applied Science and Engineering
  (Taiwan)} \bibinfo{volume}{25}, \bibinfo{pages}{215--222}.
\newblock \DOIprefix\doi{10.6180/jase.202202{\_}25(1).0022}.
%Type = Incollection
\bibitem[{Takahashi and Thepkanjana(1997)}]{Takahashi1997NegationStudy}
\bibinfo{author}{Takahashi, K.}, \bibinfo{author}{Thepkanjana, K.},
  \bibinfo{year}{1997}.
\newblock \bibinfo{title}{{Negation in Thai serial verb constructions: A
  pragmatic study}}, in: \bibinfo{booktitle}{Southeast Asian Linguistic Studies
  in Honor of Vichin Panupong}. \bibinfo{publisher}{Chulalongkorn University
  Press}, pp. \bibinfo{pages}{273--282}.
%Type = Inproceedings
\bibitem[{Tarlao et~al.(2016)Tarlao, Steele, Fernandez and
  Guastavino}]{Tarlao2016ComparingMontreal}
\bibinfo{author}{Tarlao, C.}, \bibinfo{author}{Steele, D.},
  \bibinfo{author}{Fernandez, P.}, \bibinfo{author}{Guastavino, C.},
  \bibinfo{year}{2016}.
\newblock \bibinfo{title}{{Comparing soundscape evaluations in French and
  English across three studies in Montreal}}, in:
  \bibinfo{booktitle}{Proceedings of the 45th International Congress and
  Exposition on Noise Control Engineering}, pp. \bibinfo{pages}{6855--6861}.
%Type = Article
\bibitem[{Tarlao et~al.(2021)Tarlao, Steffens and
  Guastavino}]{Tarlao2021InvestigatingModeling}
\bibinfo{author}{Tarlao, C.}, \bibinfo{author}{Steffens, J.},
  \bibinfo{author}{Guastavino, C.}, \bibinfo{year}{2021}.
\newblock \bibinfo{title}{{Investigating contextual influences on urban
  soundscape evaluations with structural equation modeling}}.
\newblock \bibinfo{journal}{Building and Environment} \bibinfo{volume}{188},
  \bibinfo{pages}{107490}.
\newblock \URLprefix \url{https://doi.org/10.1016/j.buildenv.2020.107490},
  \DOIprefix\doi{10.1016/j.buildenv.2020.107490}.
%Type = Article
\bibitem[{Tingsabadh and Abramson(1993)}]{Tingsabadh1993Thai}
\bibinfo{author}{Tingsabadh, M.R.K.}, \bibinfo{author}{Abramson, A.S.},
  \bibinfo{year}{1993}.
\newblock \bibinfo{title}{{Thai}}.
\newblock \bibinfo{journal}{Journal of the International Phonetic Association}
  \bibinfo{volume}{23}, \bibinfo{pages}{24--28}.
\newblock \URLprefix
  \url{https://www.cambridge.org/core/article/thai/E8DF7F30D887DB49BC1E9F48163CB7E6},
  \DOIprefix\doi{DOI: 10.1017/S0025100300004746}.
%Type = Article
\bibitem[{Ullman and Bentler(2012)}]{Ullman2012StructuralModeling}
\bibinfo{author}{Ullman, J.B.}, \bibinfo{author}{Bentler, P.M.},
  \bibinfo{year}{2012}.
\newblock \bibinfo{title}{{Structural equation modeling}}.
\newblock \bibinfo{journal}{Handbook of Psychology, Second Edition}
  \bibinfo{volume}{2}.
%Type = Article
\bibitem[{Wild et~al.(2005)Wild, Grove, Martin, Eremenco, McElroy,
  Verjee-Lorenz and Erikson}]{Wild2005PrinciplesAdaptation}
\bibinfo{author}{Wild, D.}, \bibinfo{author}{Grove, A.},
  \bibinfo{author}{Martin, M.}, \bibinfo{author}{Eremenco, S.},
  \bibinfo{author}{McElroy, S.}, \bibinfo{author}{Verjee-Lorenz, A.},
  \bibinfo{author}{Erikson, P.}, \bibinfo{year}{2005}.
\newblock \bibinfo{title}{{Principles of good practice for the translation and
  cultural adaptation process for patient-reported outcomes (PRO) measures:
  Report of the ISPOR Task Force for Translation and Cultural Adaptation}}.
\newblock \bibinfo{journal}{Value in Health} \bibinfo{volume}{8},
  \bibinfo{pages}{94--104}.
\newblock \URLprefix \url{http://dx.doi.org/10.1111/j.1524-4733.2005.04054.x},
  \DOIprefix\doi{10.1111/j.1524-4733.2005.04054.x}.
%Type = Misc
\bibitem[{Winskel(2013)}]{Winskel2013TheBilinguals}
\bibinfo{author}{Winskel, H.}, \bibinfo{year}{2013}.
\newblock \bibinfo{title}{{The emotional Stroop task and emotionality rating of
  negative and neutral words in late Thai-English bilinguals}}.
\newblock \URLprefix \url{http://dx.doi.org/10.1080/00207594.2013.793800},
  \DOIprefix\doi{10.1080/00207594.2013.793800}.
%Type = Article
\bibitem[{Wu et~al.(2022)Wu, Zhang and Yuan}]{Wu2022AnType}
\bibinfo{author}{Wu, C.}, \bibinfo{author}{Zhang, J.}, \bibinfo{author}{Yuan,
  Z.}, \bibinfo{year}{2022}.
\newblock \bibinfo{title}{{An ERP investigation on the second language and
  emotion perception: the role of emotion word type}}.
\newblock \bibinfo{journal}{International Journal of Bilingual Education and
  Bilingualism} \bibinfo{volume}{25}, \bibinfo{pages}{539--551}.
\newblock \URLprefix \url{https://doi.org/10.1080/13670050.2019.1703895},
  \DOIprefix\doi{10.1080/13670050.2019.1703895}.
%Type = Article
\bibitem[{Yu and Kang(2014)}]{Yu2014SoundscapeTaiwan}
\bibinfo{author}{Yu, C.J.}, \bibinfo{author}{Kang, J.}, \bibinfo{year}{2014}.
\newblock \bibinfo{title}{{Soundscape in the sustainable living environment: A
  cross-cultural comparison between the UK and Taiwan}}.
\newblock \bibinfo{journal}{Science of the Total Environment}
  \bibinfo{volume}{482-483}, \bibinfo{pages}{501--509}.
\newblock \URLprefix \url{http://dx.doi.org/10.1016/j.scitotenv.2013.10.107},
  \DOIprefix\doi{10.1016/j.scitotenv.2013.10.107}.

\end{thebibliography}

\setcounter{section}{0}
\renewcommand{\thesection}{Appendix \Alph{section}}

\setcounter{table}{0}
\renewcommand{\thetable}{\Alph{section}.\arabic{table}}

\setcounter{figure}{0}
\renewcommand{\thefigure}{\Alph{section}.\arabic{figure}}
\captionsetup{font=sf}

\onecolumn

\begin{nolinenumbers}

\section{Results of the Validation Questionnaire}
\vspace{1em}
{\centering
\noindent\includegraphics[width=\textwidth]{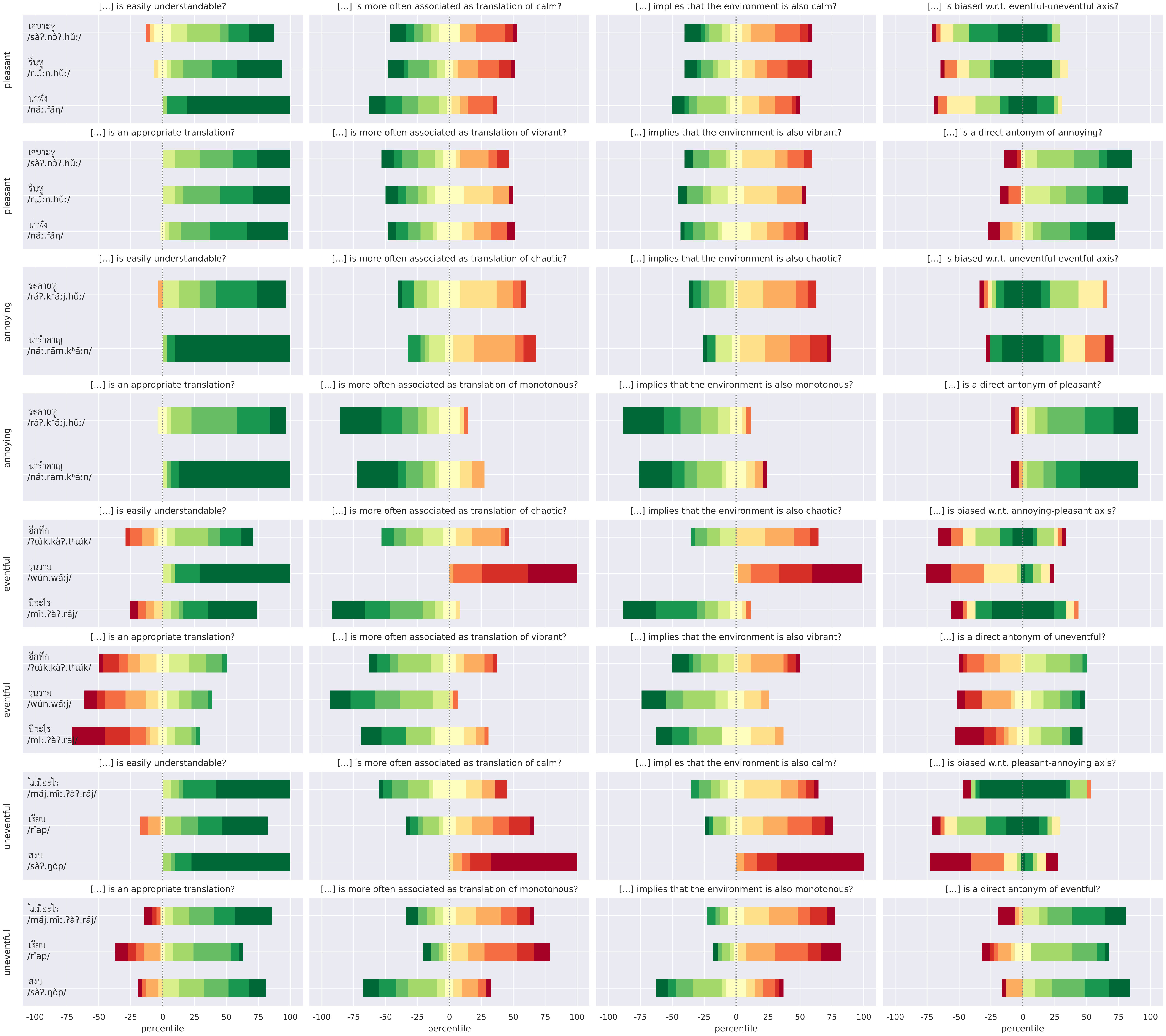}
\vskip1em
\noindent\includegraphics[width=\textwidth]{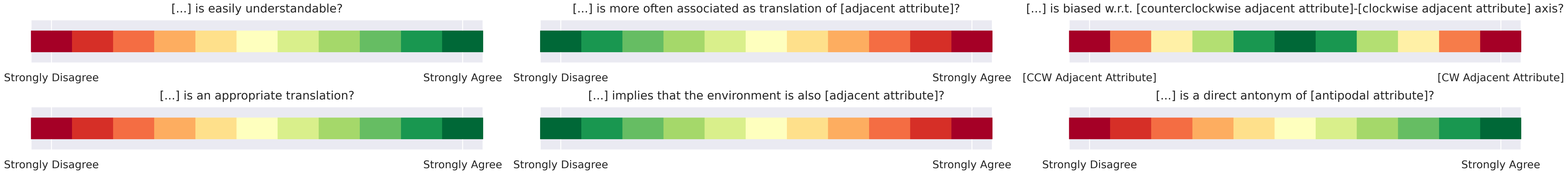}
\captionof{figure}{Validation questionnaire results for attributes on the main axes.}
\label{fig:rawmain}
}
\includegraphics[width=\textwidth]{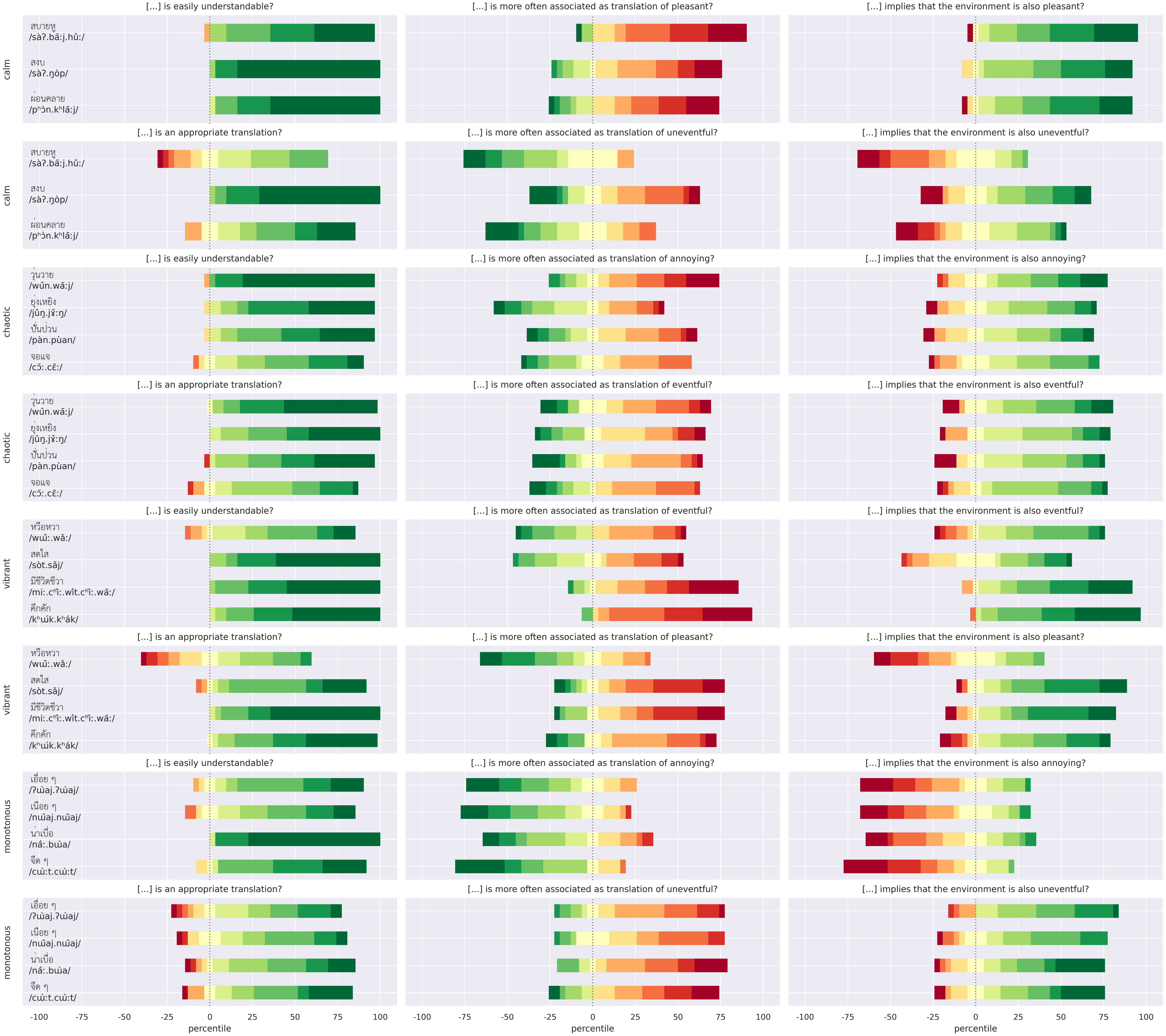}
\vskip1em
\includegraphics[width=\textwidth]{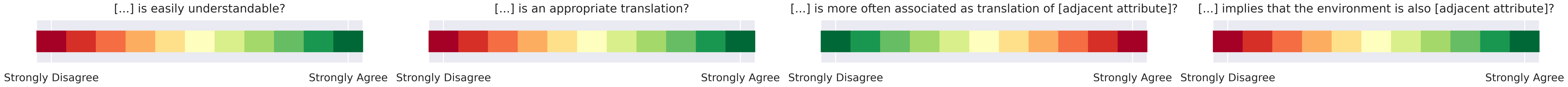}
\captionof{figure}{Validation questionnaire results for attributes on the derived axes.}
\label{fig:rawderived}

\FloatBarrier

\setcounter{table}{0}
\renewcommand{\thetable}{\Alph{section}.\arabic{table}}
\clearpage
\section{Results of Statistical Tests on the Evaluation Scores}

% \onecolumn
\begin{sffamily}
% \vskip-6pt
\stepcounter{table}
\begin{nolinenumbers}
\small
\noindent\textbf{\color{scolor}Table \thetable}\par%
\noindent{$p$-values of the Kruskal--Wallis tests and the posthoc Conover--Iman tests. Double asterisks (**) and single asterisk (*) indicate statistical significance at \SI{1}{\percent} and \SI{5}{\percent}, respectively. For Conover--Iman tests, the pairwise tests are listed in the order of increasing $p$-value, and the candidate translation with the higher average score in the respective criterion is always listed on the left of each pair.}%
% \par\vskip12pt
\end{nolinenumbers}
\small%
\begin{longtable}{llllcllrl}
% first head
    \toprule
    Eng. Attr. &
    Criterion &
    \multicolumn{5}{l}{Statistical Test} &
    $p$-value\\
    % \midrule
\endfirsthead

% head
    \toprule
    Eng. Attr. &
    Criteria &
    \multicolumn{5}{l}{Statistical Test} &
    $p$-value\\
    \midrule
\endhead

    % \cmidrule{7-9}
    \\
    \multicolumn{9}{r}{[Continued on next page]} \\
    % \midrule
\endfoot

    \bottomrule
    \\
    \multicolumn{9}{r}{[End of table]}
    \label{tab:citp}
\endlastfoot
% \midrule 
\midrule Pleasant & APPR & \multicolumn{5}{l}{Kruskal--Wallis} & 0.486 & \\*
\cmidrule{2-9}
 & UNDR & \multicolumn{5}{l}{Kruskal--Wallis} & <0.001 & ** \\
\cmidrule{3-9}
 & & \multicolumn{5}{l}{Conover--Iman} & & \\*
 & & \thainaafang & \naafang & v & \thaisanorhuu & \sanorhuu & <0.001 & ** \\*
 & & \thainaafang & \naafang & v & \thairuenhuu & \ruenhuu & <0.001 & ** \\*
 & & \thairuenhuu & \ruenhuu & v & \thaisanorhuu & \sanorhuu & 0.049 & * \\*
\cmidrule{2-9}
 & CLAR & \multicolumn{5}{l}{Kruskal--Wallis} & 0.695 & \\*
\cmidrule{2-9}
 & ANTO & \multicolumn{5}{l}{Kruskal--Wallis} & 0.957 & \\*
\cmidrule{2-9}
 & ORTH & \multicolumn{5}{l}{Kruskal--Wallis} & 0.206 & \\*
\cmidrule{2-9}
 & NCON & \multicolumn{5}{l}{Kruskal--Wallis} & 0.798 & \\*
\cmidrule{2-9}
 & IBAL & \multicolumn{5}{l}{Kruskal--Wallis} & 0.429 & \\
\midrule Annoying & APPR & \multicolumn{5}{l}{Kruskal--Wallis} & <0.001 & ** \\
\cmidrule{3-9}
 & & \multicolumn{5}{l}{Conover--Iman} & & \\*
 & & \thainaaramkaan & \naaramkaan & v & \thairakaaihuu & \rakaaihuu & <0.001 & ** \\*
\cmidrule{2-9}
 & UNDR & \multicolumn{5}{l}{Kruskal--Wallis} & <0.001 & ** \\
\cmidrule{3-9}
 & & \multicolumn{5}{l}{Conover--Iman} & & \\*
 & & \thainaaramkaan & \naaramkaan & v & \thairakaaihuu & \rakaaihuu & <0.001 & ** \\*
\cmidrule{2-9}
 & CLAR & \multicolumn{5}{l}{Kruskal--Wallis} & 0.179 & \\*
\cmidrule{2-9}
 & ANTO & \multicolumn{5}{l}{Kruskal--Wallis} & 0.085 & \\*
\cmidrule{2-9}
 & ORTH & \multicolumn{5}{l}{Kruskal--Wallis} & 0.801 & \\*
\cmidrule{2-9}
 & NCON & \multicolumn{5}{l}{Kruskal--Wallis} & 0.128 & \\*
\cmidrule{2-9}
 & IBAL & \multicolumn{5}{l}{Kruskal--Wallis} & 0.859 & \\
\midrule Eventful & APPR & \multicolumn{5}{l}{Kruskal--Wallis} & 0.030 & * \\
\cmidrule{3-9}
 & & \multicolumn{5}{l}{Conover--Iman} & & \\*
 & & \thaiuekgatuek & \uekgatuek & v & \thaimiiarai & \miiarai & 0.027 & * \\*
 & & \thaiwunwaai & \wunwaai & v & \thaimiiarai & \miiarai & 0.202 & \\*
 & & \thaiuekgatuek & \uekgatuek & v & \thaiwunwaai & \wunwaai & $\approx$1.000 & \\*
\cmidrule{2-9}
 & UNDR & \multicolumn{5}{l}{Kruskal--Wallis} & <0.001 & ** \\
\cmidrule{3-9}
 & & \multicolumn{5}{l}{Conover--Iman} & & \\*
 & & \thaiwunwaai & \wunwaai & v & \thaiuekgatuek & \uekgatuek & <0.001 & ** \\*
 & & \thaiwunwaai & \wunwaai & v & \thaimiiarai & \miiarai & 0.002 & ** \\*
 & & \thaimiiarai & \miiarai & v & \thaiuekgatuek & \uekgatuek & 0.047 & * \\*
\cmidrule{2-9}
 & CLAR & \multicolumn{5}{l}{Kruskal--Wallis} & <0.001 & ** \\
\cmidrule{3-9}
 & & \multicolumn{5}{l}{Conover--Iman} & & \\*
 & & \thaimiiarai & \miiarai & v & \thaiwunwaai & \wunwaai & <0.001 & ** \\*
 & & \thaimiiarai & \miiarai & v & \thaiuekgatuek & \uekgatuek & <0.001 & ** \\*
 & & \thaiuekgatuek & \uekgatuek & v & \thaiwunwaai & \wunwaai & 0.284 & \\*
\cmidrule{2-9}
 & ANTO & \multicolumn{5}{l}{Kruskal--Wallis} & 0.671 & \\*
\cmidrule{2-9}
 & ORTH & \multicolumn{5}{l}{Kruskal--Wallis} & <0.001 & ** \\
\cmidrule{3-9}
 & & \multicolumn{5}{l}{Conover--Iman} & & \\*
 & & \thaimiiarai & \miiarai & v & \thaiwunwaai & \wunwaai & <0.001 & ** \\*
 & & \thaiuekgatuek & \uekgatuek & v & \thaiwunwaai & \wunwaai & 0.025 & * \\*
 & & \thaimiiarai & \miiarai & v & \thaiuekgatuek & \uekgatuek & 0.055 & \\*
\cmidrule{2-9}
 & NCON & \multicolumn{5}{l}{Kruskal--Wallis} & <0.001 & ** \\
\cmidrule{3-9}
 & & \multicolumn{5}{l}{Conover--Iman} & & \\*
 & & \thaimiiarai & \miiarai & v & \thaiwunwaai & \wunwaai & <0.001 & ** \\*
 & & \thaimiiarai & \miiarai & v & \thaiuekgatuek & \uekgatuek & <0.001 & ** \\*
 & & \thaiuekgatuek & \uekgatuek & v & \thaiwunwaai & \wunwaai & 0.351 & \\*
\cmidrule{2-9}
 & IBAL & \multicolumn{5}{l}{Kruskal--Wallis} & <0.001 & ** \\
\cmidrule{3-9}
 & & \multicolumn{5}{l}{Conover--Iman} & & \\*
 & & \thaimiiarai & \miiarai & v & \thaiwunwaai & \wunwaai & <0.001 & ** \\*
 & & \thaiuekgatuek & \uekgatuek & v & \thaiwunwaai & \wunwaai & 0.002 & ** \\*
 & & \thaimiiarai & \miiarai & v & \thaiuekgatuek & \uekgatuek & 0.078 & \\
\midrule Uneventful & APPR & \multicolumn{5}{l}{Kruskal--Wallis} & 0.014 & * \\
\cmidrule{3-9}
 & & \multicolumn{5}{l}{Conover--Iman} & & \\*
 & & \thaimaimiiarai & \maimiiarai & v & \thairiiap & \riiap & 0.010 & ** \\*
 & & \thaisangop & \sangop & v & \thairiiap & \riiap & 0.261 & \\*
 & & \thaimaimiiarai & \maimiiarai & v & \thaisangop & \sangop & 0.599 & \\*
\cmidrule{2-9}
 & UNDR & \multicolumn{5}{l}{Kruskal--Wallis} & <0.001 & ** \\
\cmidrule{3-9}
 & & \multicolumn{5}{l}{Conover--Iman} & & \\*
 & & \thaisangop & \sangop & v & \thairiiap & \riiap & <0.001 & ** \\*
 & & \thaimaimiiarai & \maimiiarai & v & \thairiiap & \riiap & 0.044 & * \\*
 & & \thaisangop & \sangop & v & \thaimaimiiarai & \maimiiarai & 0.437 & \\*
\cmidrule{2-9}
 & CLAR & \multicolumn{5}{l}{Kruskal--Wallis} & 0.003 & ** \\
\cmidrule{3-9}
 & & \multicolumn{5}{l}{Conover--Iman} & & \\*
 & & \thaimaimiiarai & \maimiiarai & v & \thaisangop & \sangop & 0.003 & ** \\*
 & & \thaimaimiiarai & \maimiiarai & v & \thairiiap & \riiap & 0.051 & \\*
 & & \thairiiap & \riiap & v & \thaisangop & \sangop & $\approx$1.000 & \\*
\cmidrule{2-9}
 & ANTO & \multicolumn{5}{l}{Kruskal--Wallis} & 0.027 & * \\
\cmidrule{3-9}
 & & \multicolumn{5}{l}{Conover--Iman} & & \\*
 & & \thaisangop & \sangop & v & \thairiiap & \riiap & 0.048 & * \\*
 & & \thaimaimiiarai & \maimiiarai & v & \thairiiap & \riiap & 0.064 & \\*
 & & \thaisangop & \sangop & v & \thaimaimiiarai & \maimiiarai & $\approx$1.000 & \\*
\cmidrule{2-9}
 & ORTH & \multicolumn{5}{l}{Kruskal--Wallis} & <0.001 & ** \\
\cmidrule{3-9}
 & & \multicolumn{5}{l}{Conover--Iman} & & \\*
 & & \thaimaimiiarai & \maimiiarai & v & \thaisangop & \sangop & <0.001 & ** \\*
 & & \thairiiap & \riiap & v & \thaisangop & \sangop & <0.001 & ** \\*
 & & \thaimaimiiarai & \maimiiarai & v & \thairiiap & \riiap & 0.022 & * \\*
\cmidrule{2-9}
 & NCON & \multicolumn{5}{l}{Kruskal--Wallis} & 0.018 & * \\
\cmidrule{3-9}
 & & \multicolumn{5}{l}{Conover--Iman} & & \\*
 & & \thaimaimiiarai & \maimiiarai & v & \thaisangop & \sangop & 0.017 & * \\*
 & & \thaimaimiiarai & \maimiiarai & v & \thairiiap & \riiap & 0.138 & \\*
 & & \thairiiap & \riiap & v & \thaisangop & \sangop & $\approx$1.000 & \\*
\cmidrule{2-9}
 & IBAL & \multicolumn{5}{l}{Kruskal--Wallis} & <0.001 & ** \\
\cmidrule{3-9}
 & & \multicolumn{5}{l}{Conover--Iman} & & \\*
 & & \thairiiap & \riiap & v & \thaisangop & \sangop & <0.001 & ** \\*
 & & \thaimaimiiarai & \maimiiarai & v & \thaisangop & \sangop & <0.001 & ** \\*
 & & \thairiiap & \riiap & v & \thaimaimiiarai & \maimiiarai & $\approx$1.000 & \\
\midrule Calm & APPR & \multicolumn{5}{l}{Kruskal--Wallis} & <0.001 & ** \\
\cmidrule{3-9}
 & & \multicolumn{5}{l}{Conover--Iman} & & \\*
 & & \thaisangop & \sangop & v & \thaisabaaihuu & \sabaaihuu & <0.001 & ** \\*
 & & \thaisangop & \sangop & v & \thaipornklaai & \pornklaai & <0.001 & ** \\*
 & & \thaipornklaai & \pornklaai & v & \thaisabaaihuu & \sabaaihuu & <0.001 & ** \\*
\cmidrule{2-9}
 & UNDR & \multicolumn{5}{l}{Kruskal--Wallis} & <0.001 & ** \\
\cmidrule{3-9}
 & & \multicolumn{5}{l}{Conover--Iman} & & \\*
 & & \thaisangop & \sangop & v & \thaisabaaihuu & \sabaaihuu & <0.001 & ** \\*
 & & \thaipornklaai & \pornklaai & v & \thaisabaaihuu & \sabaaihuu & 0.021 & * \\*
 & & \thaisangop & \sangop & v & \thaipornklaai & \pornklaai & 0.279 & \\*
\cmidrule{2-9}
 & CLAR & \multicolumn{5}{l}{Kruskal--Wallis} & 0.414 & \\*
\cmidrule{2-9}
 & CONN & \multicolumn{5}{l}{Kruskal--Wallis} & 0.133 & \\*
\cmidrule{2-9}
 & IBAL & \multicolumn{5}{l}{Kruskal--Wallis} & 0.002 & ** \\
\cmidrule{3-9}
 & & \multicolumn{5}{l}{Conover--Iman} & & \\*
 & & \thaisangop & \sangop & v & \thaisabaaihuu & \sabaaihuu & 0.001 & ** \\*
 & & \thaipornklaai & \pornklaai & v & \thaisabaaihuu & \sabaaihuu & 0.072 & \\*
 & & \thaisangop & \sangop & v & \thaipornklaai & \pornklaai & 0.485 & \\
\midrule Chaotic & APPR & \multicolumn{5}{l}{Kruskal--Wallis} & <0.001 & ** \\
\cmidrule{3-9}
 & & \multicolumn{5}{l}{Conover--Iman} & & \\*
 & & \thaiwunwaai & \wunwaai & v & \thaijorjae & \jorjae & <0.001 & ** \\*
 & & \thaiyungyerng & \yungyerng & v & \thaijorjae & \jorjae & <0.001 & ** \\*
 & & \thaibpanbpuuan & \bpanbpuuan & v & \thaijorjae & \jorjae & 0.002 & ** \\*
 & & \thaiwunwaai & \wunwaai & v & \thaibpanbpuuan & \bpanbpuuan & 0.229 & \\*
 & & \thaiwunwaai & \wunwaai & v & \thaiyungyerng & \yungyerng & 0.556 & \\*
 & & \thaiyungyerng & \yungyerng & v & \thaibpanbpuuan & \bpanbpuuan & $\approx$1.000 & \\*
\cmidrule{2-9}
 & UNDR & \multicolumn{5}{l}{Kruskal--Wallis} & <0.001 & ** \\
\cmidrule{3-9}
 & & \multicolumn{5}{l}{Conover--Iman} & & \\*
 & & \thaiwunwaai & \wunwaai & v & \thaijorjae & \jorjae & <0.001 & ** \\*
 & & \thaiwunwaai & \wunwaai & v & \thaibpanbpuuan & \bpanbpuuan & <0.001 & ** \\*
 & & \thaiyungyerng & \yungyerng & v & \thaijorjae & \jorjae & 0.003 & ** \\*
 & & \thaiwunwaai & \wunwaai & v & \thaiyungyerng & \yungyerng & 0.017 & * \\*
 & & \thaibpanbpuuan & \bpanbpuuan & v & \thaijorjae & \jorjae & 0.054 & \\*
 & & \thaiyungyerng & \yungyerng & v & \thaibpanbpuuan & \bpanbpuuan & $\approx$1.000 & \\*
\cmidrule{2-9}
 & CLAR & \multicolumn{5}{l}{Kruskal--Wallis} & 0.030 & * \\
\cmidrule{3-9}
 & & \multicolumn{5}{l}{Conover--Iman} & & \\*
 & & \thaiyungyerng & \yungyerng & v & \thaiwunwaai & \wunwaai & 0.035 & * \\*
 & & \thaibpanbpuuan & \bpanbpuuan & v & \thaiwunwaai & \wunwaai & 0.133 & \\*
 & & \thaijorjae & \jorjae & v & \thaiwunwaai & \wunwaai & 0.197 & \\*
 & & \thaibpanbpuuan & \bpanbpuuan & v & \thaijorjae & \jorjae & $\approx$1.000 & \\*
 & & \thaiyungyerng & \yungyerng & v & \thaijorjae & \jorjae & $\approx$1.000 & \\*
 & & \thaiyungyerng & \yungyerng & v & \thaibpanbpuuan & \bpanbpuuan & $\approx$1.000 & \\*
\cmidrule{2-9}
 & CONN & \multicolumn{5}{l}{Kruskal--Wallis} & 0.126 & \\*
\cmidrule{2-9}
 & IBAL & \multicolumn{5}{l}{Kruskal--Wallis} & 0.771 & \\
\midrule Vibrant & APPR & \multicolumn{5}{l}{Kruskal--Wallis} & <0.001 & ** \\
\cmidrule{3-9}
 & & \multicolumn{5}{l}{Conover--Iman} & & \\*
 & & \thaimiichiiwitchiiwaa & \miichiiwitchiiwaa & v & \thaiwuewaa & \wuewaa & <0.001 & ** \\*
 & & \thaikuekkak & \kuekkak & v & \thaiwuewaa & \wuewaa & <0.001 & ** \\*
 & & \thaisotsai & \sotsai & v & \thaiwuewaa & \wuewaa & <0.001 & ** \\*
 & & \thaimiichiiwitchiiwaa & \miichiiwitchiiwaa & v & \thaisotsai & \sotsai & 0.002 & ** \\*
 & & \thaikuekkak & \kuekkak & v & \thaisotsai & \sotsai & 0.383 & \\*
 & & \thaimiichiiwitchiiwaa & \miichiiwitchiiwaa & v & \thaikuekkak & \kuekkak & 0.401 & \\*
\cmidrule{2-9}
 & UNDR & \multicolumn{5}{l}{Kruskal--Wallis} & <0.001 & ** \\
\cmidrule{3-9}
 & & \multicolumn{5}{l}{Conover--Iman} & & \\*
 & & \thaisotsai & \sotsai & v & \thaiwuewaa & \wuewaa & <0.001 & ** \\*
 & & \thaimiichiiwitchiiwaa & \miichiiwitchiiwaa & v & \thaiwuewaa & \wuewaa & <0.001 & ** \\*
 & & \thaikuekkak & \kuekkak & v & \thaiwuewaa & \wuewaa & <0.001 & ** \\*
 & & \thaimiichiiwitchiiwaa & \miichiiwitchiiwaa & v & \thaikuekkak & \kuekkak & $\approx$1.000 & \\*
 & & \thaisotsai & \sotsai & v & \thaikuekkak & \kuekkak & $\approx$1.000 & \\*
 & & \thaisotsai & \sotsai & v & \thaimiichiiwitchiiwaa & \miichiiwitchiiwaa & $\approx$1.000 & \\*
\cmidrule{2-9}
 & CLAR & \multicolumn{5}{l}{Kruskal--Wallis} & <0.001 & ** \\
\cmidrule{3-9}
 & & \multicolumn{5}{l}{Conover--Iman} & & \\*
 & & \thaiwuewaa & \wuewaa & v & \thaimiichiiwitchiiwaa & \miichiiwitchiiwaa & <0.001 & ** \\*
 & & \thaiwuewaa & \wuewaa & v & \thaikuekkak & \kuekkak & <0.001 & ** \\*
 & & \thaiwuewaa & \wuewaa & v & \thaisotsai & \sotsai & <0.001 & ** \\*
 & & \thaisotsai & \sotsai & v & \thaimiichiiwitchiiwaa & \miichiiwitchiiwaa & 0.224 & \\*
 & & \thaisotsai & \sotsai & v & \thaikuekkak & \kuekkak & 0.426 & \\*
 & & \thaikuekkak & \kuekkak & v & \thaimiichiiwitchiiwaa & \miichiiwitchiiwaa & $\approx$1.000 & \\*
\cmidrule{2-9}
 & CONN & \multicolumn{5}{l}{Kruskal--Wallis} & <0.001 & ** \\
\cmidrule{3-9}
 & & \multicolumn{5}{l}{Conover--Iman} & & \\*
 & & \thaimiichiiwitchiiwaa & \miichiiwitchiiwaa & v & \thaiwuewaa & \wuewaa & <0.001 & ** \\*
 & & \thaikuekkak & \kuekkak & v & \thaiwuewaa & \wuewaa & <0.001 & ** \\*
 & & \thaisotsai & \sotsai & v & \thaiwuewaa & \wuewaa & 0.005 & ** \\*
 & & \thaimiichiiwitchiiwaa & \miichiiwitchiiwaa & v & \thaisotsai & \sotsai & 0.080 & \\*
 & & \thaikuekkak & \kuekkak & v & \thaisotsai & \sotsai & 0.120 & \\*
 & & \thaimiichiiwitchiiwaa & \miichiiwitchiiwaa & v & \thaikuekkak & \kuekkak & $\approx$1.000 & \\*
\cmidrule{2-9}
 & IBAL & \multicolumn{5}{l}{Kruskal--Wallis} & 0.017 & * \\
\cmidrule{3-9}
 & & \multicolumn{5}{l}{Conover--Iman} & & \\*
 & & \thaimiichiiwitchiiwaa & \miichiiwitchiiwaa & v & \thaikuekkak & \kuekkak & 0.029 & * \\*
 & & \thaimiichiiwitchiiwaa & \miichiiwitchiiwaa & v & \thaiwuewaa & \wuewaa & 0.057 & \\*
 & & \thaimiichiiwitchiiwaa & \miichiiwitchiiwaa & v & \thaisotsai & \sotsai & 0.078 & \\*
 & & \thaisotsai & \sotsai & v & \thaikuekkak & \kuekkak & $\approx$1.000 & \\*
 & & \thaiwuewaa & \wuewaa & v & \thaikuekkak & \kuekkak & $\approx$1.000 & \\*
 & & \thaisotsai & \sotsai & v & \thaiwuewaa & \wuewaa & $\approx$1.000 & \\
\midrule Monotonous & APPR & \multicolumn{5}{l}{Kruskal--Wallis} & 0.425 & \\*
\cmidrule{2-9}
 & UNDR & \multicolumn{5}{l}{Kruskal--Wallis} & <0.001 & ** \\
\cmidrule{3-9}
 & & \multicolumn{5}{l}{Conover--Iman} & & \\*
 & & \thainaabuea & \naabuea & v & \thainuuainuuai & \nuuainuuai & <0.001 & ** \\*
 & & \thainaabuea & \naabuea & v & \thaiuuaiuuai & \uuaiuuai & <0.001 & ** \\*
 & & \thainaabuea & \naabuea & v & \thaijuedjued & \juedjued & <0.001 & ** \\*
 & & \thaijuedjued & \juedjued & v & \thainuuainuuai & \nuuainuuai & 0.042 & * \\*
 & & \thaijuedjued & \juedjued & v & \thaiuuaiuuai & \uuaiuuai & 0.944 & \\*
 & & \thaiuuaiuuai & \uuaiuuai & v & \thainuuainuuai & \nuuainuuai & $\approx$1.000 & \\*
\cmidrule{2-9}
 & CLAR & \multicolumn{5}{l}{Kruskal--Wallis} & 0.269 & \\*
\cmidrule{2-9}
 & CONN & \multicolumn{5}{l}{Kruskal--Wallis} & 0.608 & \\*
\cmidrule{2-9}
 & IBAL & \multicolumn{5}{l}{Kruskal--Wallis} & 0.292 & \\
\end{longtable}
\end{sffamily}
% \twocolumn

\end{nolinenumbers}

\end{document}